\documentclass[aps,prd,superscriptaddress,showpacs,tighten,nofootinbib]{revtex4-1}
\usepackage{epsfig}
\usepackage{amssymb}
\usepackage{rotate}
\usepackage{float}
\usepackage{graphicx}
\usepackage{subfig}
\usepackage{slashed}
\usepackage{setspace}
\newcommand{\ba}{\begin{array}}
\newcommand{\ea}{\end{array}}
\newcommand{\bd}{\begin{displaymath}}
\newcommand{\ed}{\end{displaymath}}
\newcommand{\be}{\begin{equation}}
\newcommand{\ee}{\end{equation}}
\newcommand{\bea}{\begin{eqnarray}}
\newcommand{\eea}{\end{eqnarray}}

\def\a{\alpha}
\def\b{\beta}
\def\g{\gamma}

\def\ve{\varepsilon}

\def\m{\mu}
\def\n{\nu}

\def\n{\nu}

\def\th13 {\theta_{13}}

\usepackage{graphicx}% Include figure files

\newcommand{\mathsym}[1]{{}}
\newcommand{\unicode}[1]{{}}

\catcode`\@=11
\def\lsim{\mathrel{\mathpalette\@versim<}}
\def\gsim{\mathrel{\mathpalette\@versim>}}
\def\@versim#1#2{\vcenter{\offinterlineskip
\ialign{$\m@th#1\hfil##\hfil$\crcr#2\crcr\sim\crcr } }}
\catcode`\@=12

\catcode`@=12
\begin{document}
\title{Non-standard interaction effect on $CP$ violation in neutrino oscillation  with super-beam}

\author{Zini Rahman}
\email{zini@ctp-jamia.res.in}
\affiliation{Centre for Theoretical Physics, Jamia Millia Islamia (Central University),  Jamia Nagar, New Delhi-110025, India}

\author{Arnab Dasgupta}
\email{arnab@ctp-jamia.res.in}
\affiliation{Centre for Theoretical Physics, Jamia Millia Islamia (Central University),  Jamia Nagar, New Delhi-110025, India}

\author{Rathin Adhikari}
\email{rathin@ctp-jamia.res.in}
\affiliation{Centre for Theoretical Physics, Jamia Millia Islamia (Central University),  Jamia Nagar, New Delhi-110025, India}

\begin{abstract}
We have studied the $CP$ violation discovery reach in neutrino oscillation experiment with superbeam in presence of non-standard interactions of neutrinos with matter for both short and long baselines. For the most important channel of oscillation ($\nu_\mu \rightarrow \nu_e$) for $CP$ violation discovery there is significant effect in the oscillation probability particularly due to NSIs' $\ve_{e\mu}$
and $\ve_{e\tau}$ for longer baseline and higher energy in comparison to other NSIs'. Interestingly for these two NSIs' (for real and higher allowed values)  there is 
possibility of better discovery reach of $CP$ violation than that with only Standard Model interactions of neutrinos with matter provided that NSI values are known. For complex NSIs' we have shown the $CP$ violation discovery reach in the plane of Dirac phase $\delta$ and
NSI phase $\phi_{ij}$. Our analysis indicates that for some values of some NSI phases total $CP$ violation may not be observable for any values of $\delta $.

\end{abstract}

\maketitle
\section{Introduction}

Among the various neutrino oscillation parameters the values of two mixing angles $\theta_{12}$ and $\theta_{23}$ have been provided by experiments with certain accuracy. The magnitude of the mass squared differences $|\Delta m^2_{31}|$ and $\Delta m^2_{21}$ are also known but the sign of $\Delta m^2_{31}$ is still unknown i.e. we still do not know exactly if neutrino masses follow normal hierarchy (NH) or inverted hierarchy (IH). Apart from that we also do not know about the $CP$ violating phase $\delta$. Earlier, the third mixing angle  $\theta_{13}$ was  not known accurately except its upper bound but recently the Double Chooz experiment \cite{double},  Daya Bay experiment  \cite{daya}, MINOS \cite{MINOS} and RENO \cite{reno} experiments have found non-zero  value of  $\sin^2 2\theta_{13}$
 at about $5 \sigma $ confidence level. The non-zero value of 
 $\theta_{13}$ is required  for the existence of $CP$ violating Dirac phase $\delta$ in the PMNS matrix.  To determine
these unknown parameters several terrestrial oscillation experiments are planned like Superbeam \cite{Itow,others5}, Neutrino factory \cite{Choubey}, Beta beam facilities \cite{Zucc,Bouch} etc.
   
In this work we consider neutrino superbeam (which mainly contains $\nu_{\mu}$ and ${\bar \nu_{\mu}}$ ) coming from CERN travelling a baseline length of 2300 Km  reaches Pyh$\ddot{\mbox{a}}$salmi (Finland) where a Liquid Argon detector is placed. We also consider another baseline of 130 Km with a superbeam source at CERN and Water Cherenkov detector placed at Fr$\acute{\mbox{e}}$jus (France). 
We do a comparative study in the discovery potentials of the $CP$ violating phase $\delta$ for these two baselines in presence of both standard and non-standard interactions. There are some earlier studies for  short and long baselines for
standard interactions \cite{raut,Dighe,others6,Agarwalla:2011hh,others4,Coloma:2012ma,Coloma:2012ut,betabeam} for neutrinos coming from superbeam.  Also there are such studies 
in presence of non-standard interactions \cite{nsi08p,Ota} before Daya Bay experiment when $\theta_{13}$ was not so precisely known. Main emphasis of our work is on exploring
$CP$ violation due to $\delta$ in the light of recent experimental findings of large value of $\theta_{13}$ and how NSI could affect the discovery reach of such $CP$ violation.

This paper is organised as follows: In section II we discuss  about NSIs' and their upper bounds. We discuss about 
the analytical expression (obtained initially in \cite{m3} for large $\theta_{13}$ ) of the probability of oscillation for $\nu_\mu \rightarrow \nu_e$ 
channel in presence of NSIs. We present graphically  how NSI effect varies for various baseline lengths $L$ and neutrino energy $E$.   In section III we discuss the experimental setup, systemetic uncertainties and errors in various parameters present
in neutrino oscillation in matter. We have also discussd the statistical approach elaborating our simulation method  in doing the numerical analysis using GLoBES \cite{globes1,globes11}.  In section IV we have presented our results of numerical analysis and
have mentioned in each subsections how we have fixed various parameters (test values). We discuss the discovery reaches of $CP$ violation due
to $\delta$  for various NSIs'. For this analysis we have considered one NSI at a time and also three NSIs' corresponding to specific generation at a time.  The off-diagonal
NSIs' in NSI matrix could be complex and the corresponding complex phases could be new sources of $CP$ violation. We have
discussed how these could affect the discovery reach of $CP$ violation due to $\delta$.  Finally in section V we conclude with remarks on NSI effects on $CP$ violation discovery for short and long baselines.

\section{NSI and its effect in neutrino ocillation probabilities for different baselines}
In addition to the Standard Model (SM) Lagrangian density we consider the following non-standard fermion-neutrino interaction in matter  defined by the Lagrangian:
\bea
\label{eq:Lang}
\mathcal{L}_{NSI}^{M}=-2\sqrt{2}G_F\ve_{\a \b}^{f P}[\bar{f}\g_{\m}Pf][\bar{\n}_{\b}\g^{\m}L\n_{\a}]  
\eea
where $P \in (L,R)$, $L=\frac{(1-\g^5)}{2}$, $R=\frac{(1+\g^5)}{2}, $ $f=e, u, d$ and $\ve_{\a \b}^{fP}$ is the deviation from SM interactions  and can be called as the non-standard interactions(NSIs). A bound can be set to these NSI parameters \cite{nsi08p, 1loop, nsi01p, nsi02p, nsi03p, nsi04p, nsi05p, nsi06p, nsi09p, nsi010p, nsi011p, sknsi} which are dependent on specific models \cite{raby, adhi} or are also model independent \cite{nsi1,Ohlsson}. These NSI parameters can be reduced to the effective parameters and can be written as:
\be
\label{eq:nsieffec}
\ve_{\a \b}=\sum_{f,P} \ve_{\a \b}^{fP}\frac{n_f}{n_e} 
\ee
where $n_f$ is the fermion number density and $n_e$ is the electron number density.
These NSIs play significant role in the context of neutrino oscillation experiments and modify the   interaction of neutrinos with matter and thus  change the oscillation probability of different flavor of neutrinos. The NSIs could be present at the source of neutrinos, during the propagation of neutrinos and also during detection of neutrinos \cite{ra}. The NSI effects are expected to be smaller at the source and
detector due to their stringent constraints \cite{nsi1,Ohlsson}. We shall consider the NSI effect during the propagation of neutrinos only. In
section IV in presenting results of our numerical analysis for one NSI at a time we shall follow the constraints as shown in the table \ref{table:bound} following references \cite{sknsi,Ohlsson,Es,Ad}.
\begin{table}[ht]
\centering % used for centering table
\begin{tabular}{|c |c |c|} % centered columns (4 columns)

\hline %inserts double horizontal lines

NSI &  Model dependent & Model indepndent \\
& bound on NSI [Reference \cite{Ohlsson}]& Bound on NSI [Reference \cite{nsi1}]\\
\hline %inserts single line
$\varepsilon_{ee}$ & $> -0.9; < 0.75$   & $ < 4.2$ \\
\hline
$|\varepsilon_{e \mu}|$ & $ \lsim 3.8 \times 10^{-4}$  & $ < 0.33$ \\
\hline
$|\varepsilon_{e \tau}|$ & $\lsim 0.25$  &  $< 3.0$ \\
\hline
$\varepsilon_{\mu \mu}$ & $  > -0.05 ;  <  0.08$  & $ < 0.068$ \\
\hline
$|\varepsilon_{\mu \tau}|$ & $ \lsim 0.25$  & $ < 0.33$ \\
\hline
$\varepsilon_{\tau \tau}$ & $ \lsim 0.4$  & $ < 21 $ \\
\hline
\end{tabular}
\caption{Strength of Non standard interaction terms used for our Analysis} % title of Table
\label{table:bound} % is used to refer this table in the text
\end{table}
There are some model independent upper bounds of different NSIs as mentioned in reference \cite{nsi1,Ohlsson} 
 which could be large under some specific conditions. 
Considering recent results from experiments in IceCube-79 and DeepCore  more  stringent bound on $\varepsilon_{\mu \mu}$, 
$|\varepsilon_{\mu \tau}|$ and $\varepsilon_{\tau \tau}$ have been obtained in \cite{Es}. However, the analysis has been done
considering two flavor only.

In vacuum, flavor eigenstates $\nu_\alpha$ may be related to  mass eigenstates of neutrinos $\nu_i$ as
\be
\vert\nu_\alpha>=\sum_{i}  U_{\alpha i}\vert\nu_i>
\; ;\;\quad 
\qquad i=1, 2, 3,
\ee
where $U$ is PMNS matrix \cite{pmns,pmns1} which depends on three mixing angles $\theta_{12}$, $\theta_{23}$ and $\theta_{13}$ and one 
$CP$ violating phase $\delta$. Although two more Majorana phases could be present in $U$ but are not relevant for neurino oscillation experiments.
The Hamiltonian due to standard ($H_{SM} $) and non-standard interactions ($H_{NSI}$) of neutrinos interacting with matter during propagation can be written in the flavor basis as:
\begin{eqnarray}
\label{eq:hamil}
H=H_{SM} + H_{NSI}
\end{eqnarray}
where
\newline
\begin{eqnarray}
\label{eq:H}
H_{SM} = \frac{\Delta m^2_{31}}{2E}\left[U\pmatrix{0 & 0 & 0 \cr 0 & \alpha & 0 \cr
 0
 & 0 & 1} U^{\dag}+\pmatrix{A & 0 & 0 \cr 0 & 0 & 0 \cr
 0
 & 0 & 0} \right], \nonumber \\
 \end{eqnarray}
 \begin{eqnarray}
 \label{eq:H1}
&&H_{\text{NSI}} =
A \pmatrix{\varepsilon_{e e} & \varepsilon_{e \mu} & \varepsilon_{e \tau} \cr
\varepsilon_{e \mu}^* & \varepsilon_{\mu \mu} & \varepsilon_{\mu \tau} \cr
\varepsilon_{e \tau}^* & \varepsilon_{\mu \tau}^* & \varepsilon_{\tau \tau}} \nonumber \\
\end{eqnarray}
 In equations (\ref{eq:H}) and (\ref{eq:H1}) 
\bea
\label{eq:matternsi}
A = \frac{2E\sqrt{2}G_{F}n_{e}}{\Delta m_{31}^2} ; \; 
\alpha = \frac{\Delta m^2_{21}}{\Delta m^2_{31}} ; \;
\Delta m^2_{i j} = m^2_i - m^2_j  \nonumber \\
\eea
where $m_i$ is the mass of the $i$-th neutrino and $A$ is considered due to the interaction of neutrinos with
matter in SM,  $G_{F}$ is the Fermi constant and $n_{e}$ is the electron number density of matter . $\varepsilon_{ee}$, $\varepsilon_{e\mu}$ , $\varepsilon_{e\tau}$, $\varepsilon_{\mu\mu}$, $\varepsilon_{\mu\tau}$
 and $\varepsilon_{\tau\tau}$  are considered due to the non-standard interaction (NSIs) of neutrinos with  matter. In equation (\ref{eq:H1}), ($\; ^{*} \;$)  
 denotes complex conjugation. 
In general, the NSIs - $\ve_{e\mu}$, $\ve_{e\tau}$ and $\ve_{\mu\tau}$ could be complex and later in the expressions of probability of oscillation these are expressed as $\ve_{ij}=|\ve_{ij}|e^{i \phi_{ij}}$.

\subsection{Oscillation Probability}

For  understanding qualitatively the NSI effect in finding the  discovery reach of  $CP$ violation  we present below the oscillation probability $P_{\nu_{\mu}\rightarrow \nu_{e}}$ .
Oscillation channels $\nu_{\mu} \rightarrow \nu_e$ and $\bar{\nu}_{\mu} \rightarrow \bar{\nu}_e$  are particularly sensitive to $CP$ violation.  In our numerical analysis however, we have
considered other channels also like $\nu_{\mu} \rightarrow \nu_{\mu}$ and $\bar{\nu}_{\mu} \rightarrow \bar{\nu}_{\mu}$. To get the expression  for $P_{\nu_{\mu} \rightarrow \nu_e}$ 
the $\sin\theta_{13} \sim \mathcal{O}(\sqrt{\a})$ has been considered which is very near to the best fit value of recent reactor experiments.  Considering NSI parameters $\ve_{\a\b} $ of the order of $\a$  one obtains upto order $\a^2$ \cite{m3}
\bea
\label{eq:prob}
P_{\nu_{\mu}\rightarrow \nu_e} &=& P^{SM}_{\nu_{\mu}\rightarrow \nu_e} + P^{NSI}_{\nu_{\mu}\rightarrow \nu_e}   
\eea
where
\begin{widetext}
\bea
\label{eq:prob1}
&&P^{SM}_{\nu_{\mu}\rightarrow \nu_e} = 4\sin \frac{(A-1)\Delta m^2_{31}L}{4E}\frac{s^2_{13}s^2_{23}}{(A-1)^4}\bigg(((A-1)^2-(1+A)^2s^2_{13})\sin\frac{(A-1)\Delta m^2_{31}L}{4E}\nonumber \\
&+&A(A-1)\frac{\Delta m^2_{31}L}{E}s^2_{13}\cos\frac{(A-1)\Delta m^2_{31}L}{4E}\bigg)
+ \frac{\a^2c^2_{23}}{A^2}\sin^2 2\theta_{12}\sin^2\frac{\Delta m^2_{31}AL}{4E}\nonumber \\
&+&\frac{\a s^2_{12}s^2_{13}s^2_{23}}{(A-1)^3}\bigg(\frac{(A-1)\Delta m^2_{31}L}{E}\sin \frac{(A-1)\Delta m^2_{31}L}{2E}-8A\sin^2 \frac{(A-1)\Delta m^2_{31}L}{4E}\bigg)\nonumber \\
&+& \frac{\a s_{13}s_{2 \times 12}s_{2 \times 23}}{A(A-1)}\bigg(2\cos\bigg(\delta + \frac{\Delta m^2_{31}L}{4E}\bigg)\sin \frac{(A-1)\Delta m^2_{31}L}{4E}\sin\frac{A\Delta m^2_{31}L}{4E}\bigg) 
\eea
\end{widetext}
\begin{widetext}
\bea
\label{eq:pro2}
P^{NSI}_{\nu_{\mu}\rightarrow \nu_e} &&= \frac{4|a_2|s_{2\times 23}s_{13}}{A(A-1)}\sin \frac{A\Delta m^2_{31}L}{4E}\sin\frac{(A-1)\Delta m^2_{31}L}{4E}\cos \big(\delta + \frac{\Delta m^2_{31}L}{4E}+\phi_{a_2}\big)\nonumber \\
&+& \frac{4|a_3|s^2_{23}}{(A-1)^2}\sin^2 \frac{(A-1)\Delta m^2_{31}L}{4E}(|a_3|+2\cos(\delta + \phi_{a_3})s_{13})\nonumber \\
&+&\frac{s^2_{13}s^2_{23}(|a_5|-|a_1|)}{(A-1)^3E}\bigg(8E\sin^2 \frac{(A-1)\Delta m^2_{31}L}{4E}-(A-1)\Delta m^2_{31}L\sin \frac{(A-1)\Delta m^2_{31}L}{2E}\bigg)\nonumber \\
&+&\frac{4 |a_2|c_{23}}{(A-1)A^2}\sin \frac{A\Delta m^2_{31}L}{4E}\bigg((A-1)c_{23}\sin \frac{A\Delta m^2_{31}L}{4E}(|a_2|+\a\cos \phi_{a_2}\sin 2\theta_{12})\bigg)\nonumber \\
&-&\frac{4|a_2| |a_3|\sin 2\theta_{23}}{A(A-1)}\cos\bigg[\frac{\Delta m^2_{31}L}{4E}+\phi _{a_2}-\phi_{a_3}\bigg]\sin \frac{(1-A)\Delta m^2_{31}L}{4E}\sin \frac{A\Delta m^2_{31}L}{4E}\nonumber \\
&+& \frac{4|a_3|s_{23}}{(A-1)^2A}\sin \frac{(A-1)\Delta m^2_{31}L}{4E}(A-1)\a c_{23}\cos\bigg[\frac{\Delta m^2_{31}L}{4E}+\phi _{a_3}\bigg]\sin \frac{A\Delta m^2_{31}L}{4E}\sin 2\theta_{12}\nonumber \\
&+&\frac{|a_4|s^2_{13}\sin 2 \theta_{23}}{(A-1)^2A}\sin \frac{(A-1)\Delta m^2_{31}L}{4E}\bigg(-4A\cos\frac{A\Delta m^2_{31}L}{4E} \cos \phi_{a_4}\sin \frac{\Delta m^2_{31}L}{4E}\nonumber\\
&+&4\sin \frac{A\Delta m^2_{31}L}{4E}\bigg(\cos\frac{\Delta m^2_{31}L}{4E}\cos \phi_{a_4}-(A-1)\sin\frac{\Delta m^2_{31}L}{4E}\sin\phi_{a_4}\bigg)\bigg) 
\eea
\end{widetext}
where
\bea
&&a_1 = A \ve_{e e} \nonumber \\
&&|a_2|e^{i \phi_{a_2}} = A \bigg(e^{i\phi_{e\mu}}|\ve_{e\mu}|c_{23}-e^{i\phi_{e\tau}}|\ve_{e\tau}|s_{23}\bigg) \nonumber \\ 
&&|a_3|e^{i \phi_{a_3}} = A \bigg(e^{i\phi_{e\tau}}|\ve_{e\tau}|c_{23}+e^{i\phi_{e\mu}}|\ve_{e\mu}|s_{23}\bigg) \nonumber \\
&&|a_4|e^{i \phi_{a_4}} = A \bigg(|\ve_{\mu \tau}|e^{i\phi_{\mu \tau}} - 2 |\ve_{\mu \tau}|s^2_{23} + (\ve_{\mu \mu}-\ve_{\tau \tau})c_{23}s_{23}\bigg) \nonumber \\
&&a_5 = A \bigg(  \ve_{\tau \tau} c^2_{23} +  \ve_{\mu \mu} s^2_{23} +  |\ve_{\mu \tau}| \cos \phi_{\mu \tau}s_{2 \times 23}\bigg) \nonumber \\
%&&|a_6| = A \Big{(}|\epsilon_{\mu \mu}| c_{23}^2-2 |\epsilon_{\mu \tau}| c_{23} \cos (\phi_{\mu \tau}) s_{23}+|\epsilon_{\tau \tau}| s_{23}^2\Big{)} \nonumber \\
\eea

and
\bea
\phi_{a_2}&=&\tan^{-1}\left[\frac{\vert\ve_{e \mu}\vert c_{23}  \sin \phi_{e\mu} -\vert\ve_{e \tau}\vert s_{23}  \sin \phi_{e \tau}}{\vert\ve_{e \mu}\vert c_{23}  \cos \phi_{e\mu}]-\vert\ve_{e \tau}\vert \cos \phi_{e \tau}] s_{23} }\right]\nonumber \\
\phi_{a_3}&=&\tan^{-1}\left[\frac{\vert\ve_{e\mu}\vert s_{23}  \sin \phi_{e \mu}+\vert\ve_{e \tau}\vert c_{23} \sin \phi_{e \tau}}{\vert\ve_{e \tau}\vert c_{23}  \cos \phi_{e \tau} +\vert\ve_{e\mu}\vert \cos \phi_{e \mu} s_{23} }\right] \nonumber \\
\phi_{a_4} &=& \tan^{-1}\left( \frac{|\varepsilon_{\mu \tau}|\sin(\phi_{\mu \tau})}{|\varepsilon_{\mu \tau}|c_{2\times 23}\cos(\phi_{\mu \tau})+(\varepsilon_{\mu \mu} - \varepsilon_{\tau \tau})c_{23} s_{23}}\right) \nonumber \\
\eea
where $s_{ij}=\sin \theta_{ij}$, $c_{ij}=\cos \theta_{ij}$, $s_{2 \times ij}=\sin 2\theta_{ij}$, $c_{2 \times ij}=\cos 2\theta_{ij}$.

One can relate the oscillation  probabilities for antineutrinos to those probabilities given for neutrinos above  
by the following relation:
%%%%%%%%%%%%%%%%%%%%%%%%%%%%%%%%%%%%%%%%%%%%%%%%%%%%%%%%%%%%%%%%%%%%%%%%%%%%%%%%%%%%
\begin{equation}
P_{\bar{\alpha}\bar{\beta}}= P_{\alpha \beta}(\delta_{CP} \rightarrow 
-\delta_{CP},\; {A} \rightarrow 
- { A}).
\label{probsd}
\end{equation}

%%%%%%%%%%%%%%%%%%%%%%%%%%%%%%%%%%%%%%%%%%%%%%%%%%%%%%%%%%%%%%%%%%%%%%%%%%%%%%%%%%%

In addition, we also have to replace  $\ve_{\alpha\beta}$
with their
complex conjugates, in order to deduce the oscillation probability for the antineutrino, 
if one considers non-standard interaction during propagation.

 For only SM interactions, (i.e  $\ve_{\a\b} \rightarrow 0$) in above expressions of oscillation probabilities one finds that for both short and long baseline the 
$\delta$ dependence occurs at order of $\a^{3/2}$.  However, for large values of $A$ for longer baseline the matter effect is
more than short baseline in the $\delta$ independent part of the oscillation probability. For this reason the discovery reach of $CP$ violation is better in short baseline than that in long baseline.  
However, when NSIs are also taken into account one can see that for longer baseline further $\delta $ dependence in
  $P_{\nu_{\mu} \rightarrow \nu_e}$ could occur at the order of $\a^{3/2}$ through  $a_3$ containing terms in (\ref{eq:prob}) for NSIs of the order of $\a$. We have checked that for slightly higher 
NSIs of the order of $ \sqrt{\a}$ using  perturbation method the same $\delta$ dependent terms in  $P_{\nu_{\mu} \rightarrow \nu_e}$ appears with  $a_3$ in the oscillation probability for long baseline as given in
(\ref{eq:prob}) and this slightly higher NSI makes these terms at the order of $\a$ which could compete with the $\delta$ independent but matter dependent part  in 
 $P_{\nu_{\mu} \rightarrow \nu_e}$ for long baseline as that  is also at the order of $\a$. This improvement of $\delta$ dependent part over independent part
for long baseline does not happen for short baseline as  $a_3$ is small for short baseline due to small value of $A$.  So presence of  NSIs  in $a_3$ improves the discovery reach of $CP$ violation for longer baseline
in comparison to the short baseline.  As  $a_3$ contains NSIs like $\ve_{e\mu}$
and $\ve_{e\tau}$ it is expected that in presence of these NSIs the long baseline could provide a better discovery reach for $CP$ violation. However, as the upper bound of $\ve_{e\mu}$ is somewhat smaller one may expect $\ve_{e\tau}$ will have significant effect in the discovery reach of $CP$ violation. In fact, our numerical analysis also shows this feature. Effect of other NSI's appearing through other $a_i$  in
the expression of oscillation probability are not coupled to $\delta$ dependent term in the oscillation probability expression and as such in general it is expected that their effect will be lesser in the $CP$ violation discovery search.

\begin{figure}[H]
\centering
\begin{tabular}{cc}
\includegraphics[width=0.5\textwidth]{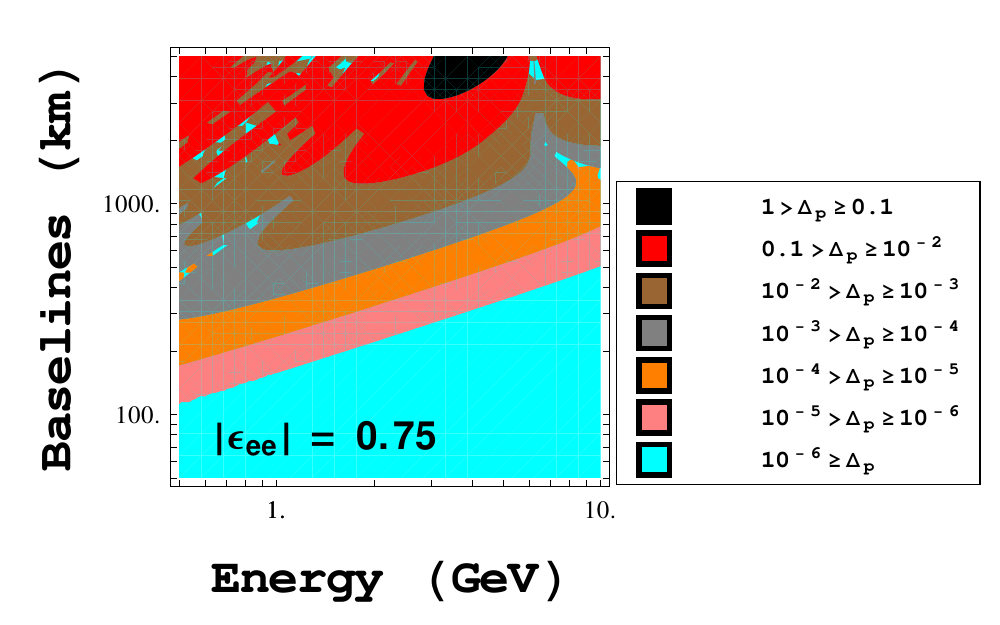}&
\includegraphics[width=0.3\textwidth]{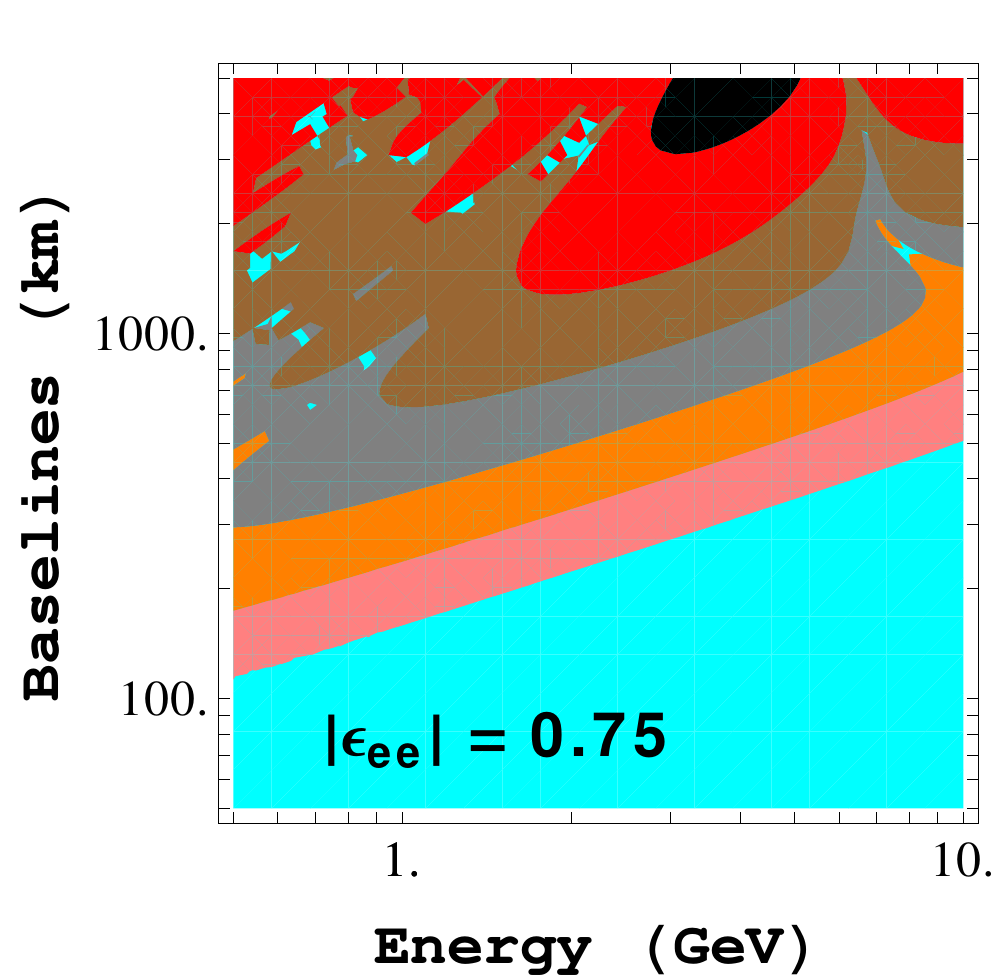}\\
\includegraphics[width=0.3\textwidth]{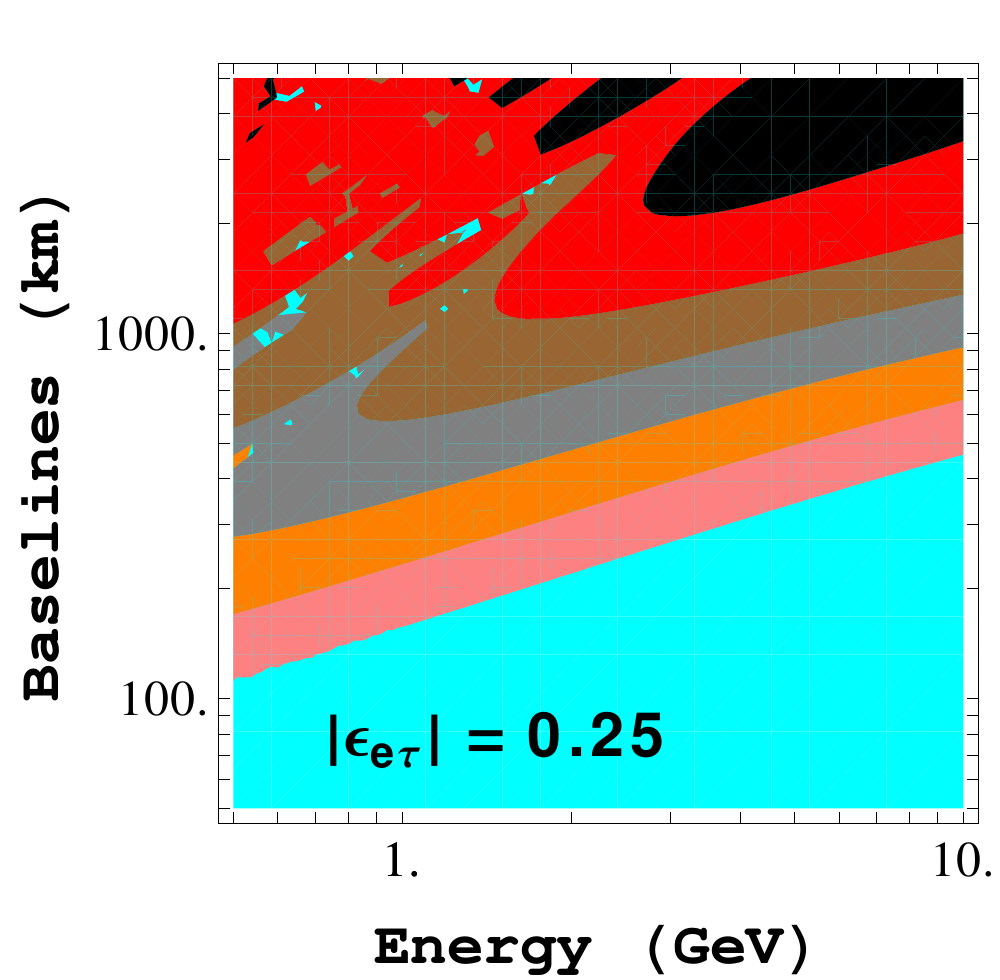}&
\includegraphics[width=0.3\textwidth]{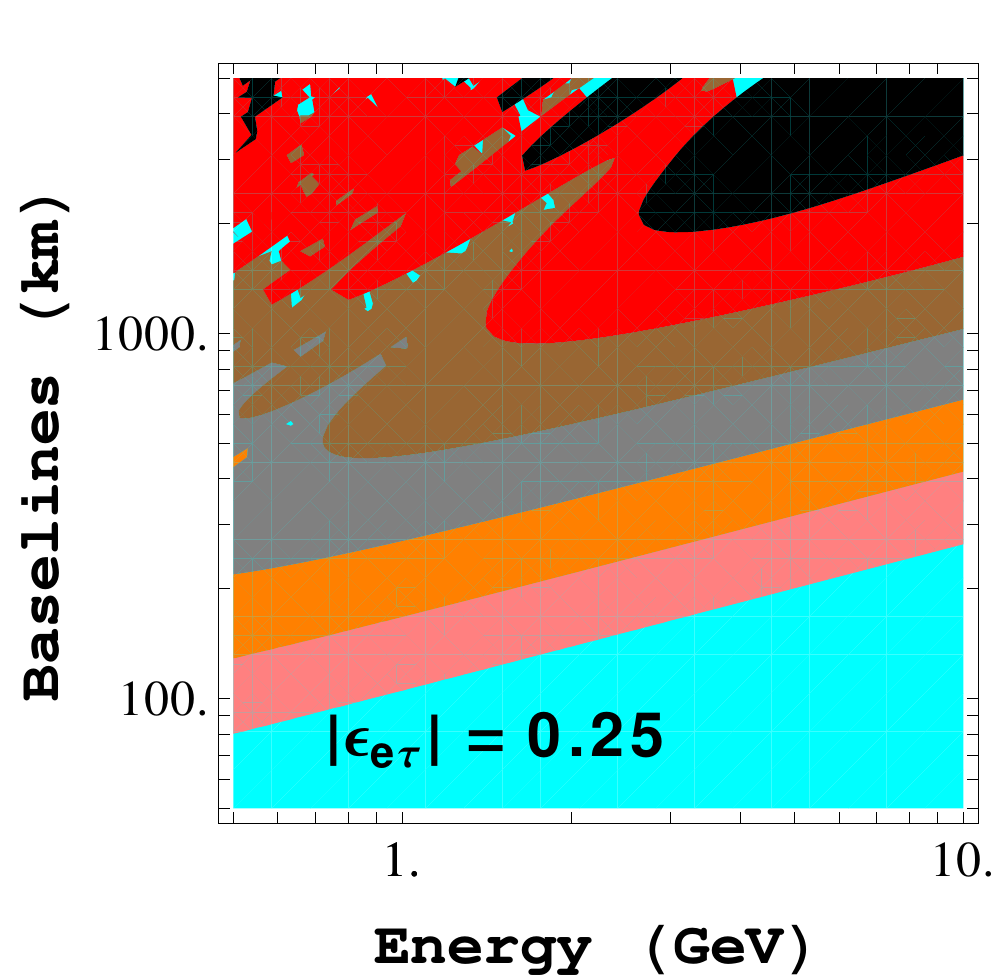}
\end{tabular}
\caption[] {{\small NSI effect $\Delta_p = | P^{SM+NSI}_{\nu_{\mu}\rightarrow \nu_e}-P^{SM}_{\nu_{\mu}\rightarrow \nu_e} |$ for different baseline length $L$ and neutrino energy $E$. Upper left panel for $\ve_{ee} = 0.75 , \delta =0$, Upper right panel for $\ve_{ee} = 0.75 , \delta =\pi/4$, Lower left panel for $\ve_{e\tau} = 0.25 , \delta =0$, Lower right panel for $\ve_{e\tau} = 0.25 , \delta =\pi/4$.}}
\label{fig:probEL1}
\end{figure}

\begin{figure}[H]
\centering
\begin{tabular}{cc}
\includegraphics[width=0.3\textwidth]{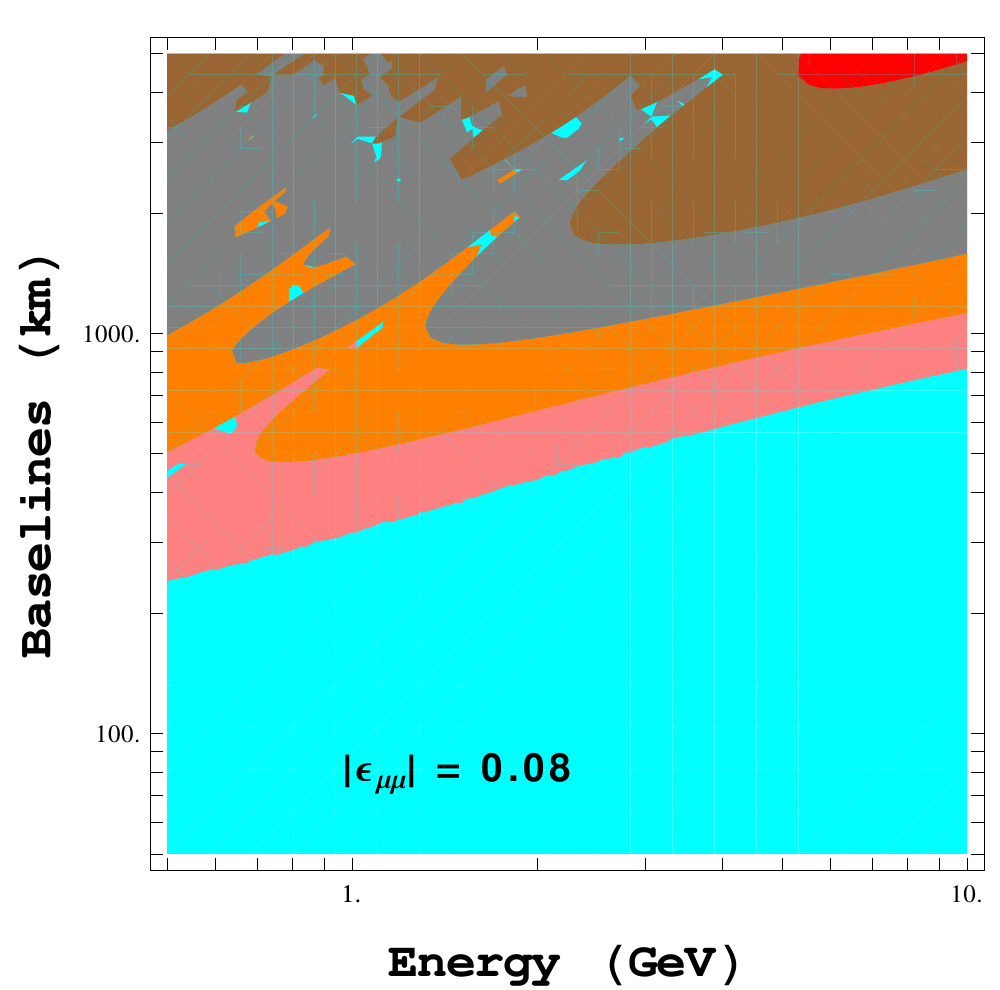}&
\includegraphics[width=0.3\textwidth]{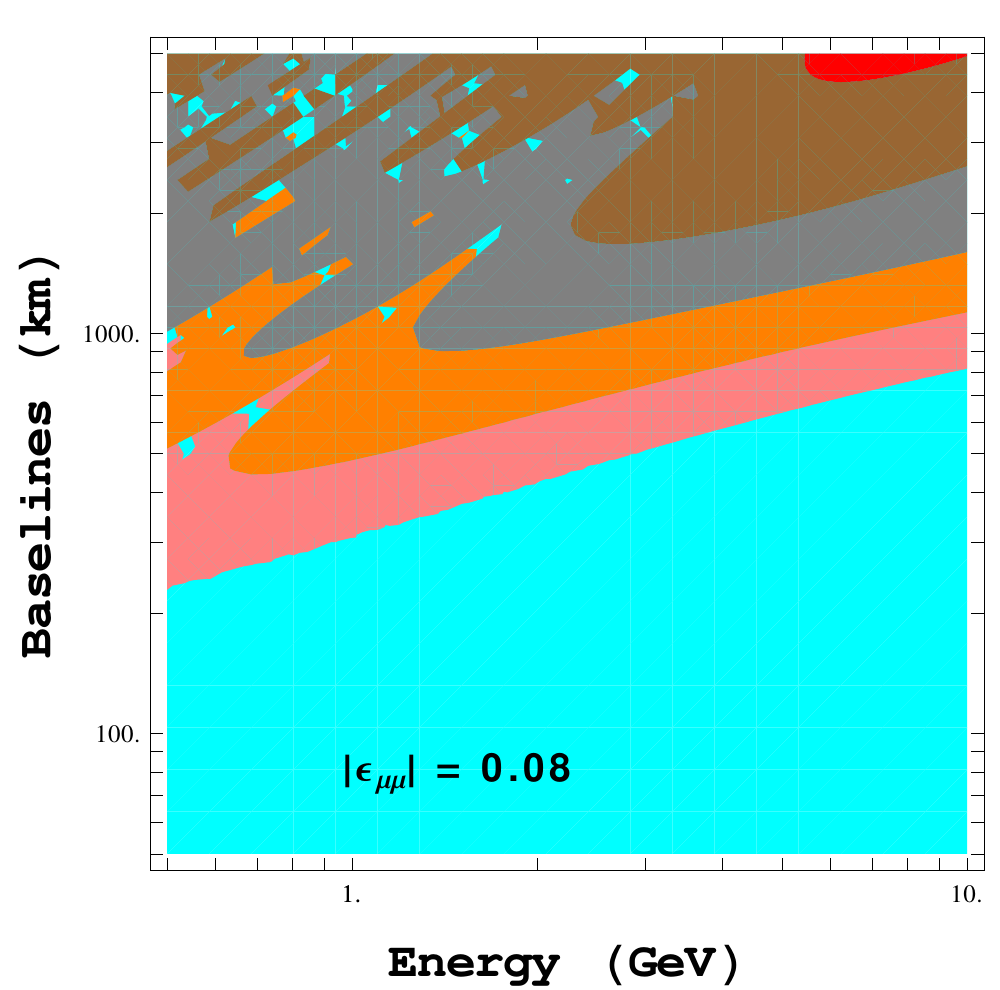}\\
\includegraphics[width=0.3\textwidth]{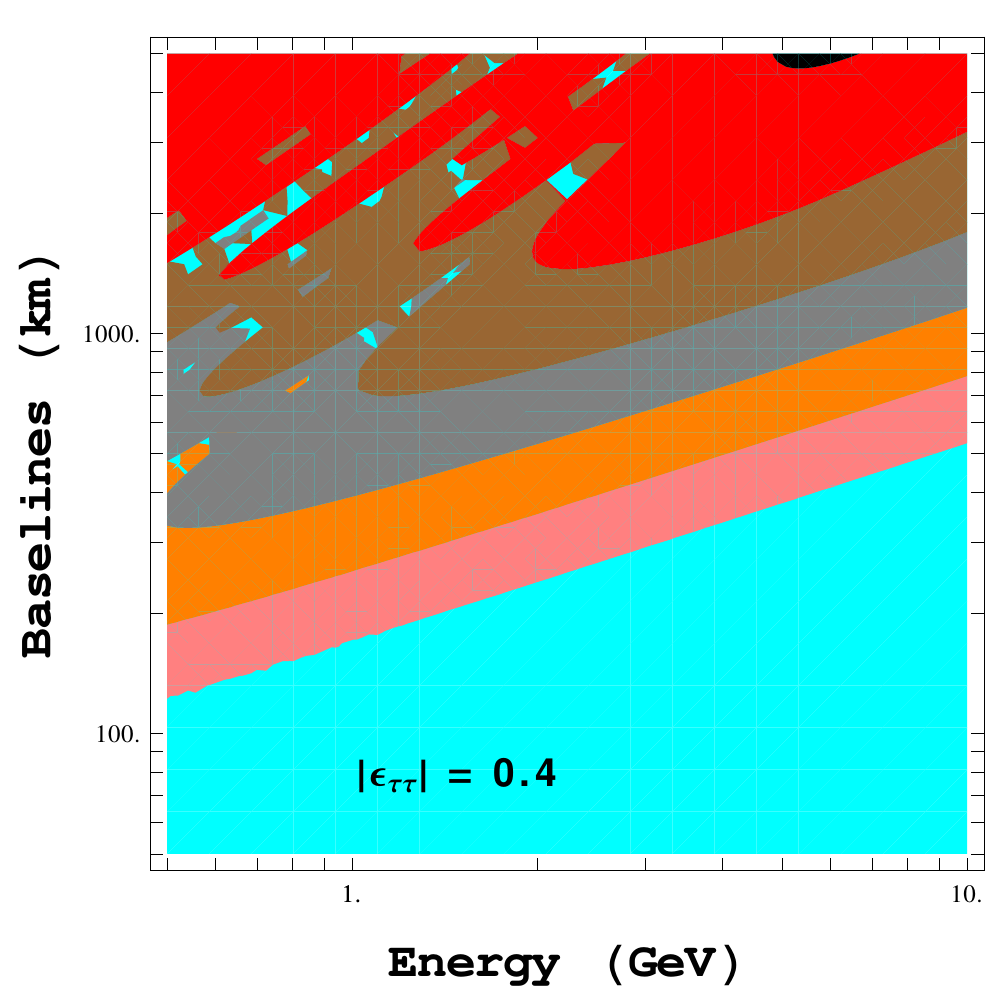}&
\includegraphics[width=0.3\textwidth]{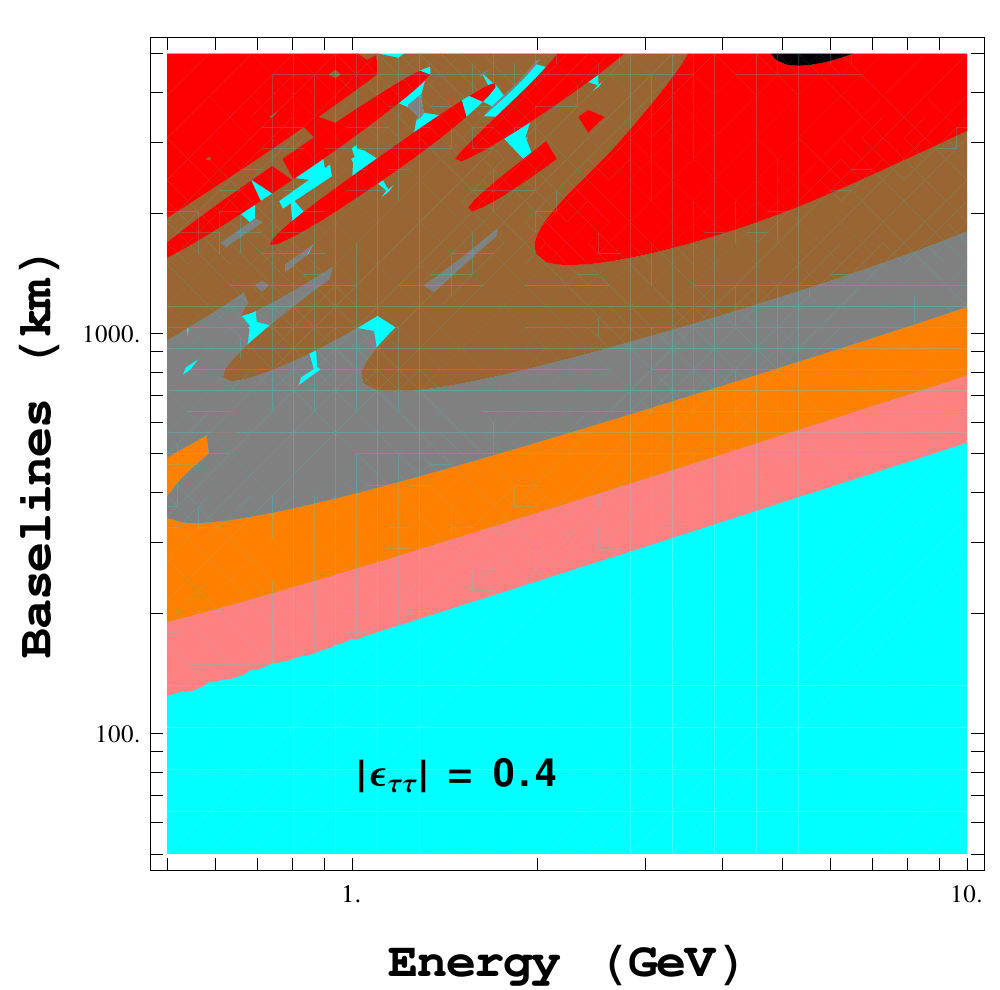}
\end{tabular}
\caption[] {{\small NSI effect $\Delta_p = | P^{SM+NSI}_{\nu_{\mu}\rightarrow \nu_e}-P^{SM}_{\nu_{\mu}\rightarrow \nu_e} |$ for different baseline length $L$ and neutrino energy $E$. Upper left panel for $\ve_{\mu\mu} = 0.08 , \delta =0$, Upper right panel for $\ve_{\mu\mu} = 0.08 , \delta =\pi/4$, Lower left panel for $\ve_{\tau\tau} = 0.4 , \delta =0$, Lower right panel for $\ve_{\tau\tau} = 0.4 , \delta =\pi/4$.}}
\label{fig:probEL2}
\end{figure}

\begin{figure}[H]
\centering
\begin{tabular}{cc}
\includegraphics[width=0.3\textwidth]{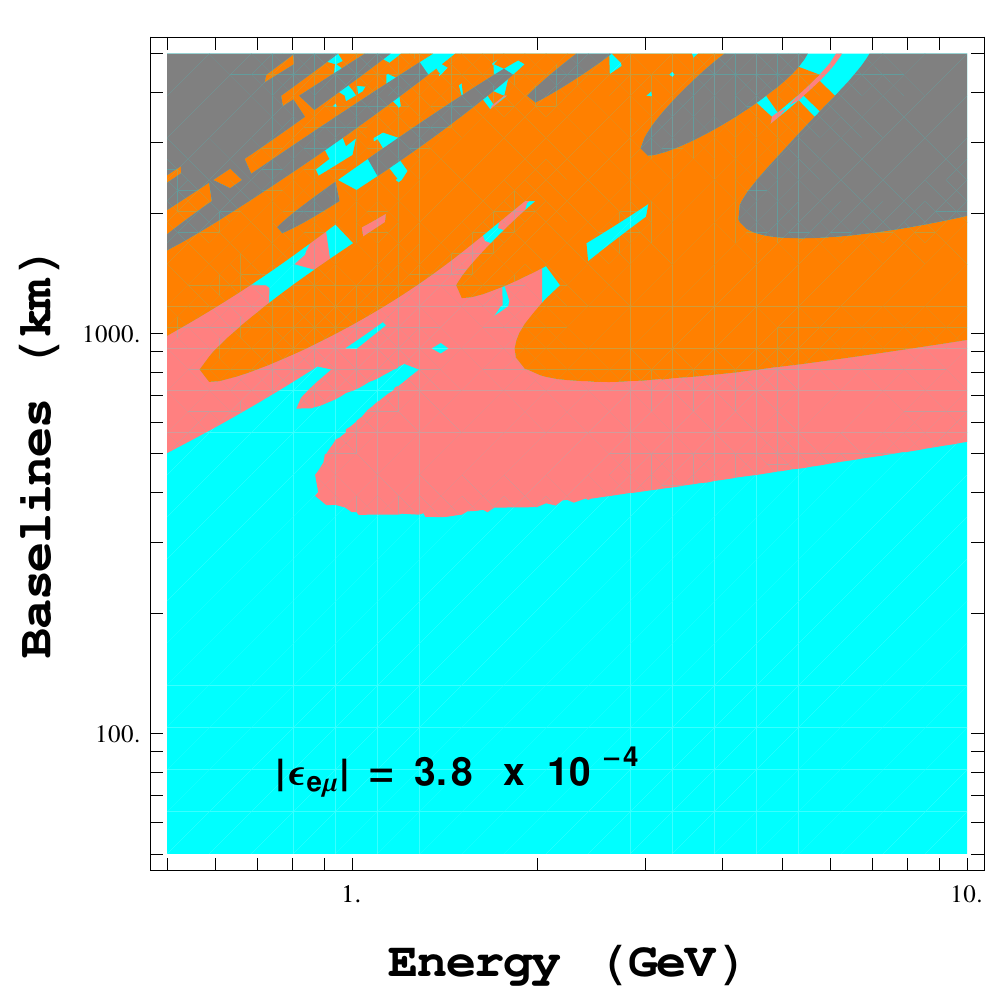}&
\includegraphics[width=0.3\textwidth]{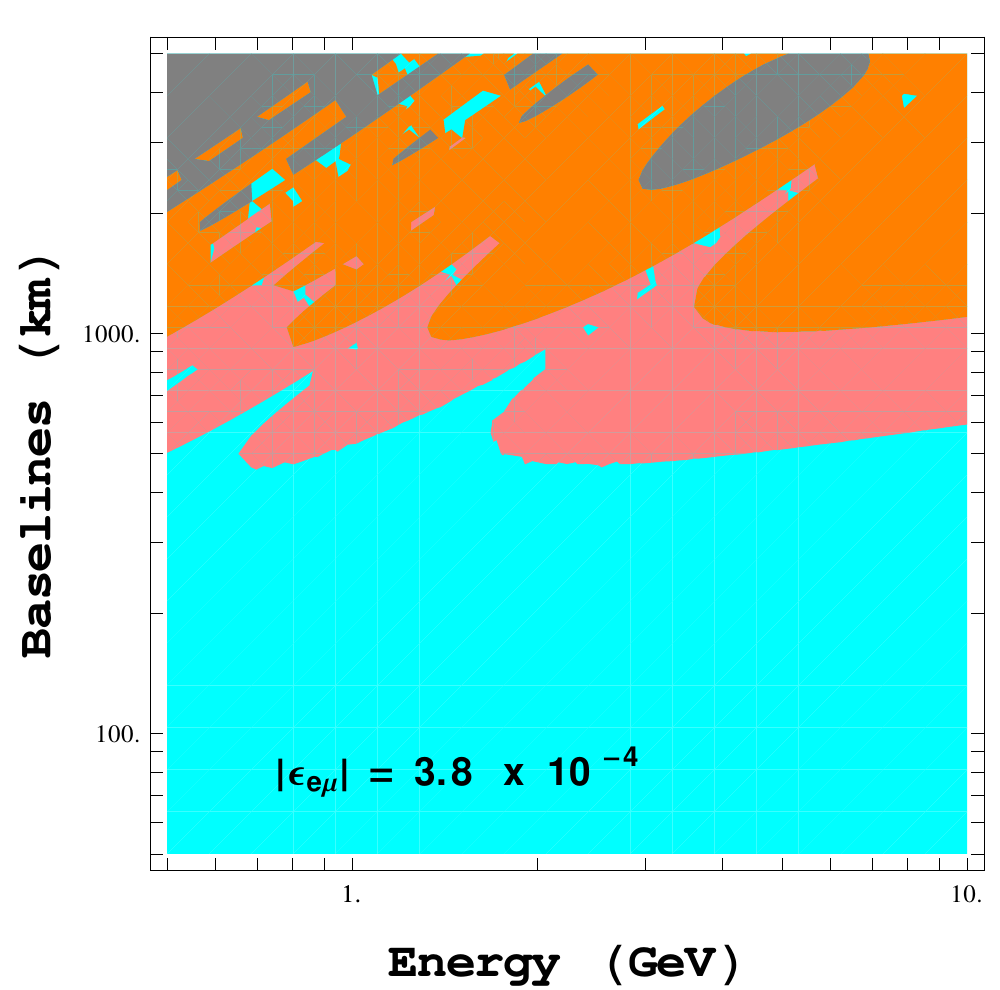}\\
\includegraphics[width=0.3\textwidth]{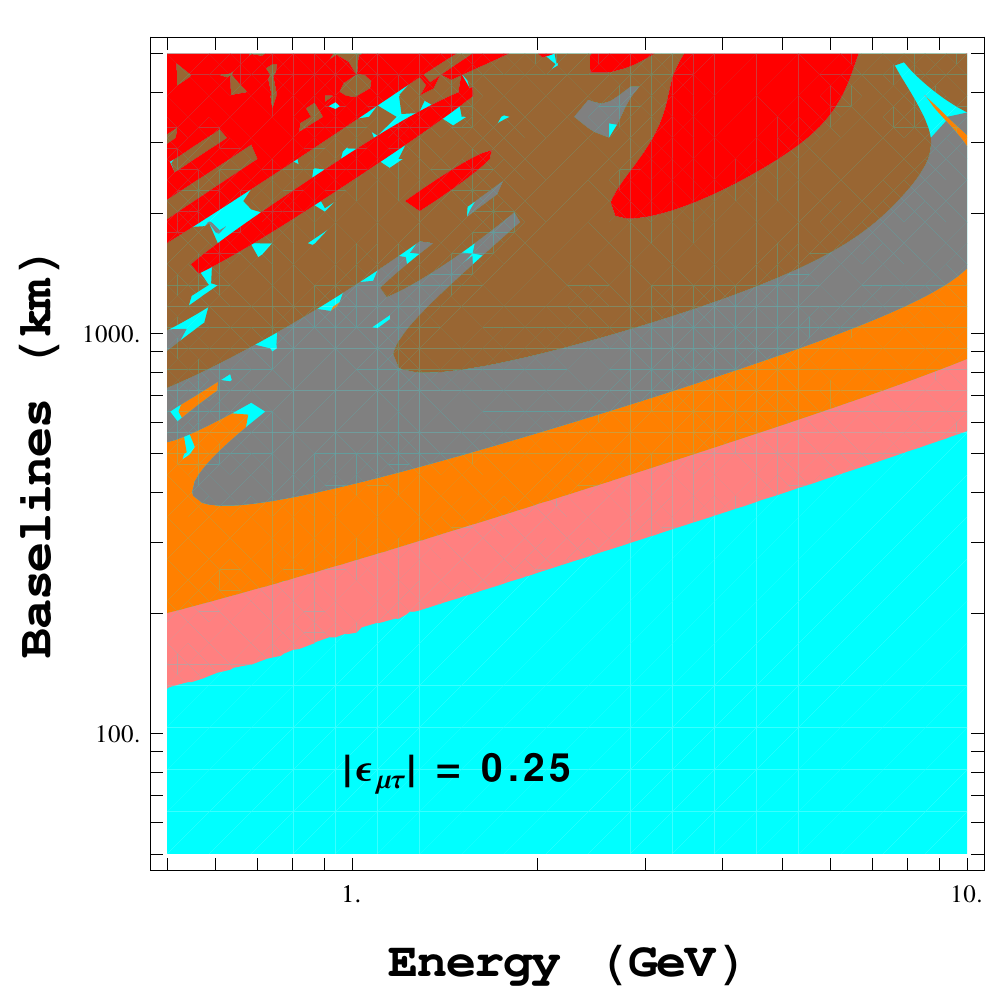}&
\includegraphics[width=0.3\textwidth]{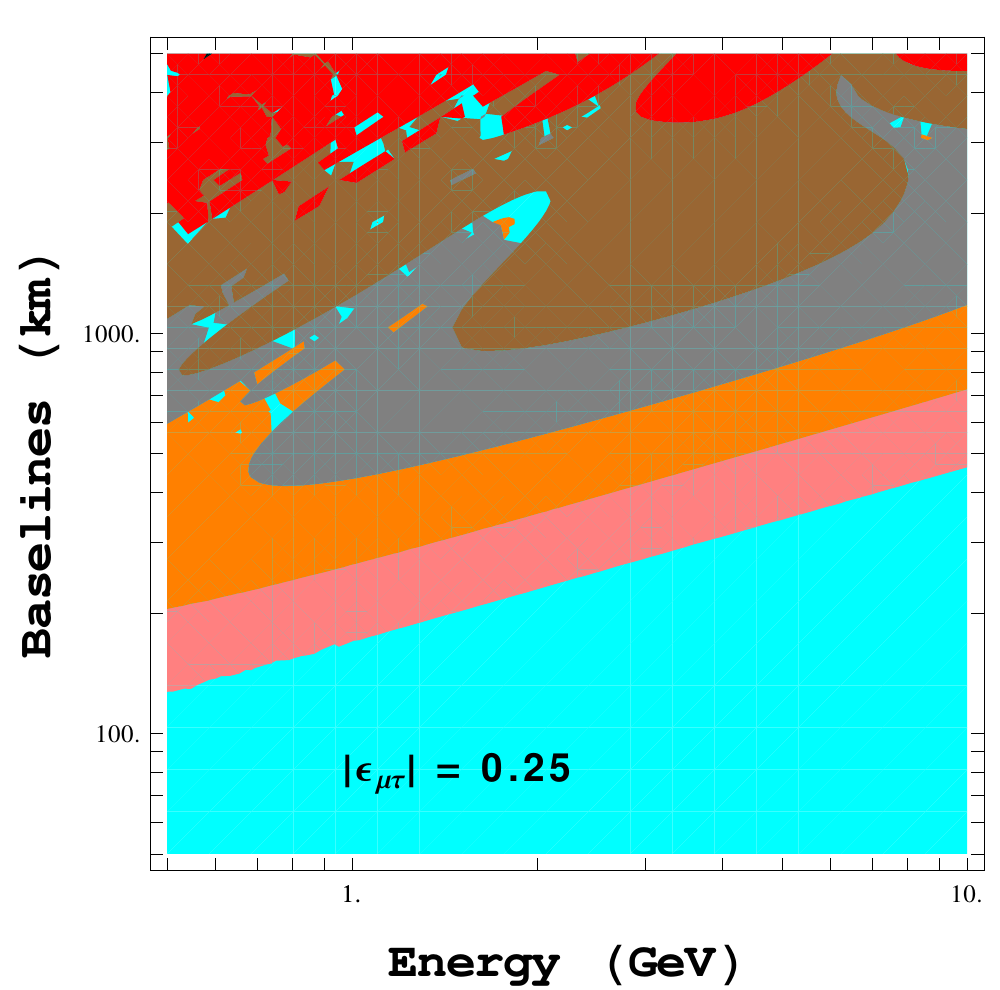}
\end{tabular}
\caption[] {{\small NSI effect $\Delta_p = | P^{SM+NSI}_{\nu_{\mu}\rightarrow \nu_e}-P^{SM}_{\nu_{\mu}\rightarrow \nu_e} |$ for different baseline length $L$ and neutrino energy $E$. Upper left panel for $\ve_{e\mu} = 3.8 \times 10^{-4} , \delta =0$, Upper right panel for $\ve_{e\mu} = 3.8 \times 10^{-4} , \delta =\pi/4$, Lower left panel for $\ve_{\mu\tau} = 0.25 , \delta =0$, Lower right panel for $\ve_{\mu\tau} = 0.25 , \delta =\pi/4$.}}
\label{fig:probEL3}
\end{figure}
To show the NSI effect $\Delta_p$ in the oscillation channel $\nu_{\mu}\rightarrow \nu_e$ at different baseline length $L$ and different neutrino energy $E$ we define it as
$\Delta_p = | P^{NSI}_{\nu_{\mu}\rightarrow \nu_e}-P^{SM}_{\nu_{\mu}\rightarrow \nu_e} |$. In figures \ref{fig:probEL1}, \ref{fig:probEL2} and \ref{fig:probEL3}  for different ranges of $\Delta_p$ different shadings have been considered as shown in figures. The NSI effects are shown in  figures with NSIs' at their upper bound values for $\delta =0$ and $\delta = \pi/4$.
For different $\delta$ values result does not change significantly. 
In figure \ref{fig:probEL1} for $\ve_{ee} =0.75$ the NSI effect $\Delta_p$ is found to be strong particularly for baselines from 3500 Km to 5000 Km and energy in the range of 3 to 5 GeV for both $\delta$ values.
For $\ve_{e\tau} =0.25$ the NSI effect $\Delta_p$ is found to be strongest particularly for baselines from 2000 Km to 5000 Km and energy in the range of 1.5 to 10 GeV for both $\delta$ values. 
In figure \ref{fig:probEL2} for $\ve_{\mu\mu} = 0.08$ it is found to be strongest particularly for baselines from 4000 Km to 5000 Km and energy in the range of 6 to 10 GeV for both $\delta$ values and the effect is relatively lesser than the earlier mentioned two NSIs'. 
For $\ve_{\tau\tau} = 0.4$ the NSI effect $\Delta_p$ is found to be  strongest particularly for baselines around 5000 Km and energy in the range of 5 to 6 GeV for both $\delta$ values. 
In figure \ref{fig:probEL3} for $\ve_{e\mu} = 3.8 \times 10^{-4}$ the NSI effect $\Delta_p$ is found to be not so strong and even lesser than $\ve_{\mu\mu}$ case. 
For $\ve_{\mu\tau} = 0.25$ the NSI effect $\Delta_p$ is found to be  slightly stronger in comparison to $\ve_{e\mu} $ particularly for baselines from 1700 Km to 5000 Km and energy in the range of 0.5 to 1.5  GeV and 3 to 6.5 GeV for both $\delta$ values. 
For the baselines which we have chosen for studying discovery reach of $CP$ violation 
the one with 2300 Km has in general medium NSI effect and the one with 130 Km has very small NSI effect.

\section{Experimental setup, systemetic uncertainties and errors of various parameters}
In this work for the numerical simulation we consider two set-ups:
(a) A Superbeam setup which originates in CERN and reaches a 100 kt Liquid Argon detector placed at a distance of 2300 Km at Pyh$\ddot{\mbox a}$salmi (Finland) (b) A Superbeam setup originating in CERN and reaching a 500 Kt Water Cherenkov detector 
 \cite{Agostino:2012fd} placed at a distance of 130 Km at Fr$\acute{\mbox{e}}$jus (France).

\begin{table}[ht]

\centering % used for centering table

\begin{tabular}{|c |c |c |} % centered columns (4 columns)

\hline %inserts double horizontal lines
Oscillation Parameters & Central Values & Error (\%) \\
\hline
$\Delta m^2_{12}$ & $7.5 \times 10^{-5}$ $eV^2$& 3  \\
\hline
$\Delta m^2_{31}$ & $2.5 \times 10^{-3}$ $eV^2$& 3  \\
\hline
$\sin^2\theta_{12}$ & $0.31$ & 5  \\
\hline
$ \theta_{23}$ & $38.3^\circ$ & 8  \\
\hline
$\sin^2 2\theta_{13}$ & 0.094  & 5  \\
\hline
\end{tabular}
\caption{Central values of the oscillation parameters with errors} % title of Table

\label{table:mix} % is used to refer this table in the text

\end{table}

For both the set-ups  
following ref \cite{1307.0807} we consider the true values of the neutrino oscillation parameters as shown in table
\ref{table:mix} and have taken into account the errors in the statistical analysis.
For earth matter
density the PREM profile \cite{prem} has been considered. Also we have considered an error of $2\%$ on matter density profile.

The flux considered for set-up (a) has  mean energy $\sim 5$ GeV, which are computed for 50 GeV protons and $3.10^{21}$ protons 
on target per year. For our analysis the beam power has been considered of about 0.8 MW per year and the time period has been taken to be 5 years each for neutrinos and anti-neutrinos.
We consider the same flux as in \cite{others6}. The detector characteristics for set-up (a) is given in table \ref{table:charac} \cite{Adams:2013qkq}.
The correlation between the visible energy of background NC events and the neutrino energy is
implemented by migration matrices which has been  provided by L. Whitehead \cite{white}. 

\begin{table}[ht]

\centering % used for centering table

\begin{tabular}{|c |c |c |c|} % centered columns (4 columns)

\hline
\hline %inserts double horizontal lines
 Signal Studies& $\nu_e$ CC appearance Studies & $\nu_\mu$ CC Disppearance Studies  \\
\hline
Signal efficiency &  80\% & 85\%  \\
\hline
$\nu_\mu$ NC mis-identification rate (Background) & 1\%  & 0.5\% \\
\hline
$\nu_\mu$ CC mis-identification rate (Background) & 1\% & 0\% \\
\hline 
Signal Normalization error & 5\% & 10\% \\
\hline
Background Normalization Error & 15\% & 20\% \\
\hline
\multicolumn{3}{|c|}{Neutrino Energy Resolution} \\
\hline
\hline
$\nu_e$ CC energy resolution & \multicolumn{2}{|c|}{0.15$\sqrt{E}$} \\
\hline
$\nu_\mu$ CC energy resolution & \multicolumn{2}{|c|}{0.2$\sqrt{E}$} \\
\hline
$E_{\nu_\mu}$ scale uncertainity &  \multicolumn{2}{|c|}{2\%} \\
\hline
$E_{\nu_e}$ scale uncertainity &  \multicolumn{2}{|c|}{0.01\%} \\
\hline
\end{tabular}
\caption{Detector characteristics for set-up (a)} % title of Table

\label{table:charac} % is used to refer this table in the text

\end{table}

The flux considered for set-up (b) has  mean energy $\sim 0.3$ GeV, which are computed for 3.5 GeV protons and $10^{23}$ protons on target per year. For our analysis the beam power has been considered of about 4 MW per year and the time period has been taken to be 2 years for neutrinos and 4 years for anti-neutrinos.
We consider the same flux as in \cite{others5,splflux}.
In the case of set-up(b)  the efficiencies for the signal and background are 
included in the migration matrices based on \cite{Agostino:2012fd} except for the channels $\nu_\mu$ disappearance, $\bar{\nu}_\mu$ 
disappearance and $\nu_\mu$ (NC) which are $64\%$, $81\%$ and  $11.7\%$ efficiencies respectively. We have considered systematic uncertainties of $2\%$ on signal and background channels.

\begin{table}[ht]

\centering % used for centering table

\begin{tabular}{|c |c |c |c|} % centered columns (4 columns)

\hline
\hline %inserts double horizontal lines

NSI & Strength of NSI & 130 (Km) & 2300 (Km) \\ [0.5ex] % inserts table 

%heading
%\hline
\hline % inserts single horizontal line

$\varepsilon$'s & 0 & 4358 & 397 \\ % inserting body of the table
\hline
$\varepsilon_{ee}$ & 0.375 & 4392 & 446 \\
 & 0.75 & 4425 & 484 \\
\hline
$\varepsilon_{e\mu}$ & 1.9$\times 10^{-4}$ & 4359 & 397 \\
 &$3.8\times10^{-4}$ & 4360 & 397 \\
\hline
$\varepsilon_{e\tau}$ & 0.125 & 4398 & 627 \\
 & 0.25	& 4440 & 960 \\
\hline 
$\varepsilon_{\mu \tau}$ & $0.125$& 4344 & 375 \\
 &$0.25$ & 4329 & 355 \\
\hline
$\varepsilon_{\mu \mu}$ & 0.04& 4359 & 403 \\
 & 0.08& 4360 & 409 \\
\hline
$\varepsilon_{\tau \tau}$ & 0.2 & 4337 & 338 \\
 & 0.4 & 4316 & 287 \\

\hline %inserts single line

\end{tabular}
\caption{Total number of events for ${\nu_{\mu}\rightarrow\nu_e}$ oscillation for no NSI and for different NSI's} % title of Table

\label{table:event} % is used to refer this table in the text

\end{table}
In table \ref{table:event} we have shown the expected number of events for two baselines for no NSIs' and also for
medium and upper most allowed values of different NSIs' and have considered the central values of various parameters
as shown in table \ref{table:mix} and matter densities are 2.7 gm/cc and 3.1378 gm/cc for 130 Km and 2300 Km respectively. Except for NSIs' $\ve_{\mu\tau}$, $\ve_{e\mu}$ and $\ve_{\tau\tau}$ for other NSIs' the expected number of
events for ${\nu_{\mu}\rightarrow\nu_e}$ oscillation channel are found to be more than the number of events for no NSIs'.
Particularly for $\ve_{e\tau}$ there is significant increase in the number of events with respect to no NSI case. This matches with our discussion in section II. 
  
In doing the whole analysis we have used GLoBES software \cite{globes1,globes11} and for taking into account NSIs'  we have followed the method described in \cite{globes11} and modified the source file "probability.c" in GLoBES appropriately
and inserted the NSI's in the subroutine where the hamiltonian for matter interaction is defined. Then we have included the new probability program as instructed by the manual of GLoBES.

In performing the $\chi^2$ analysis the observable channels that we have considered in the $\chi^2$ analysis are 
$\nu_{\mu}\rightarrow \nu_e$, $\bar{\nu}_{\mu}\rightarrow \bar{\nu}_e$, $\nu_{\mu}\rightarrow \nu_{\mu}$ and $\bar{\nu}_{\mu}\rightarrow \bar{\nu}_{\mu} $. We have considered a Poissonian $\chi^2$ with priors over the parameters $\theta_{12}, \theta_{13}, \theta_{23}, \delta_{CP}, \Delta m^2_{21}, \Delta m^2_{31}, \rho$. In the next section while presenting the results for various discovery reaches we have mentioned how test values of some parameters have been fixed.

\section{Results}

In this work we have done a comparative study for the two experimental set-ups (a) and (b) in finding the discovery reach of the $CP$ violation due to  only Dirac phase $\delta$ in PMNS matrix and also due to this Dirac phase as well as one of the NSI phases (corresponding three NSIs' in the off-diagonal elements of NSI matrix) for both hierarchies. 

\subsection{Discovery reach of $CP$ violation due to $\delta$ for real NSIs'}
To compare with the $CP$ violation discovery reach in presence of NSI with that in absence of NSIs' we present below the earlier known results on $CP$ violation discovery reach for SM interactions of neutrinos with matter. We have fixed $\delta^{test}$ to 0 and $\pi$ and have marginalized over hierarchy for every  $\delta^{true}$ value. 

\begin{figure}[H]
\centering
\begin{tabular}{cc}
\includegraphics[width=0.5\textwidth]{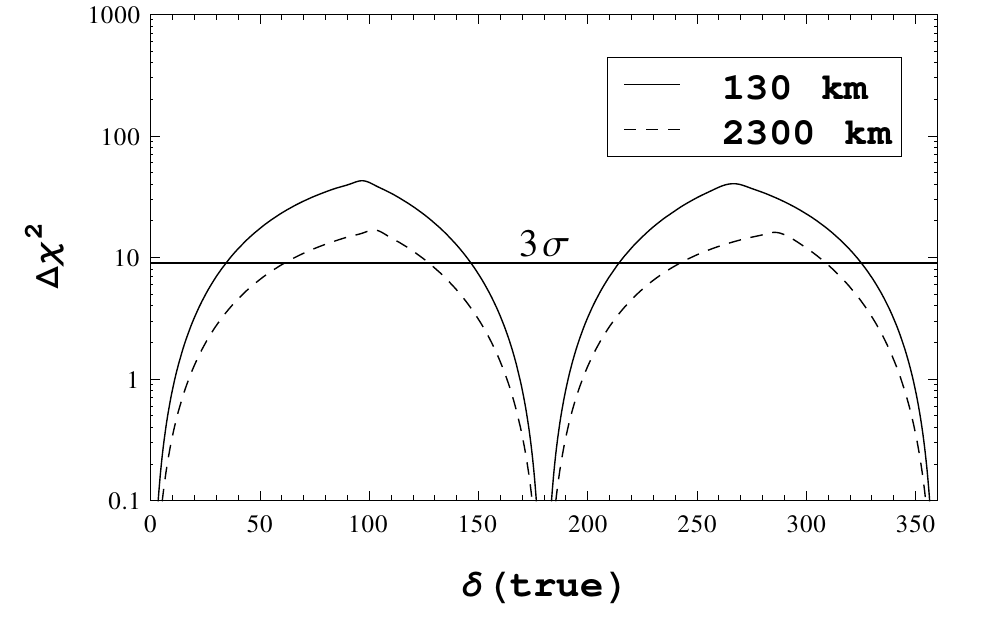}&
\includegraphics[width=0.5\textwidth]{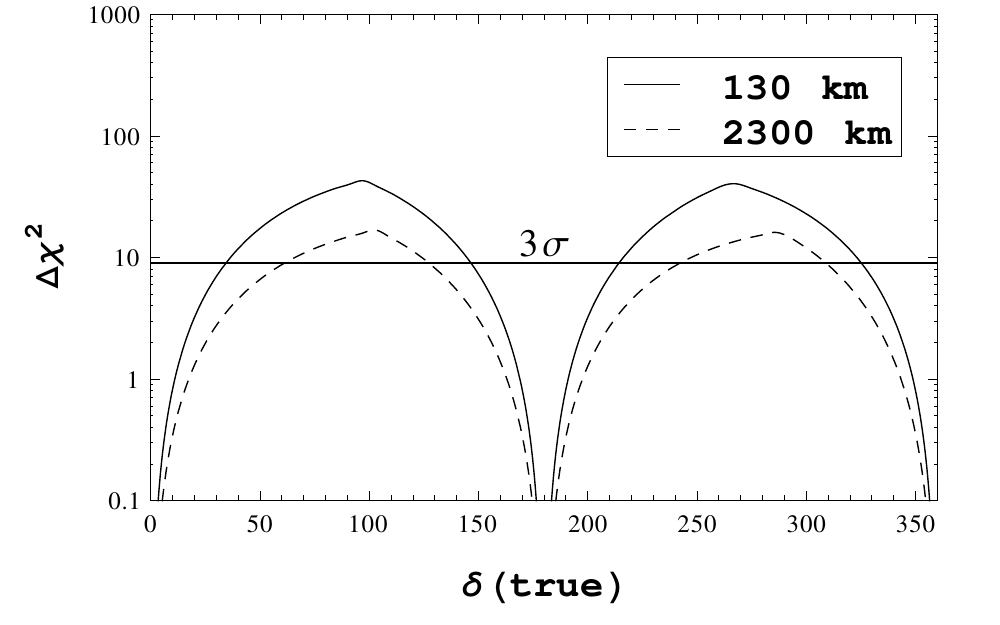}
\end{tabular}
\caption[] {{\small Discovery reach of CP violation due to $\delta$ for two different baselines 130 Km and 2300 Km considering only SM interactions. The  left panel is for NH and the right panel is for IH.}}
\label{fig:delta1}
\end{figure}

\begin{figure}[H]
\centering
\begin{tabular}{cc}
\includegraphics[width=0.5\textwidth]{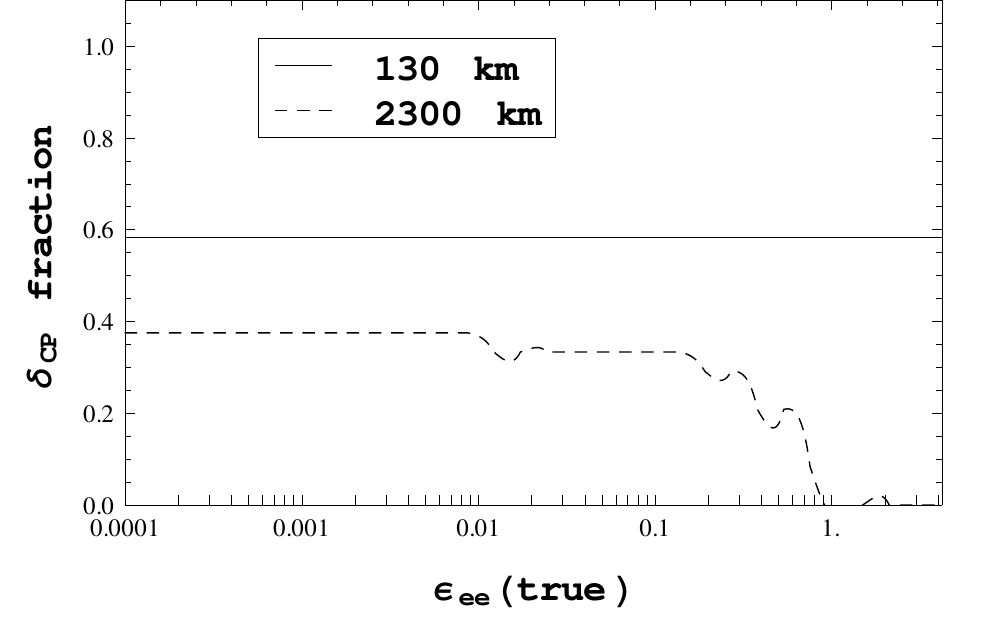}&
\includegraphics[width=0.5\textwidth]{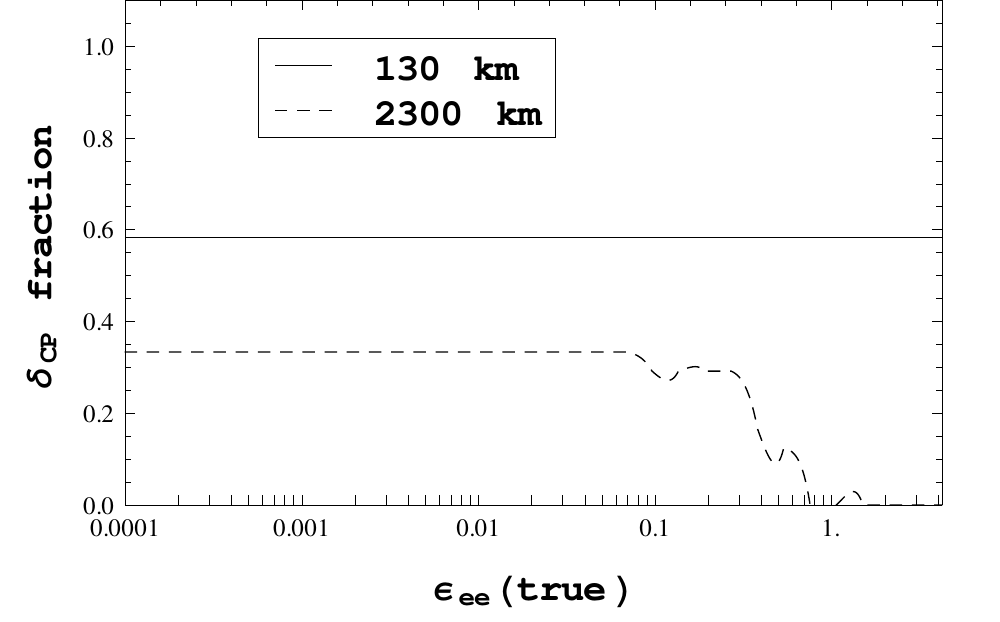}\\
\includegraphics[width=0.5\textwidth]{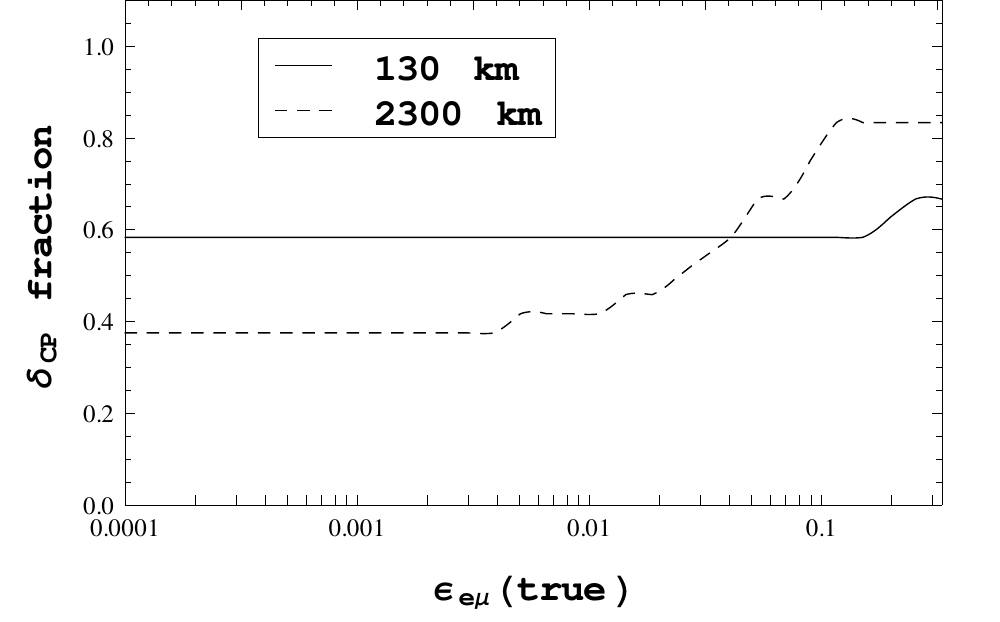}&
\includegraphics[width=0.5\textwidth]{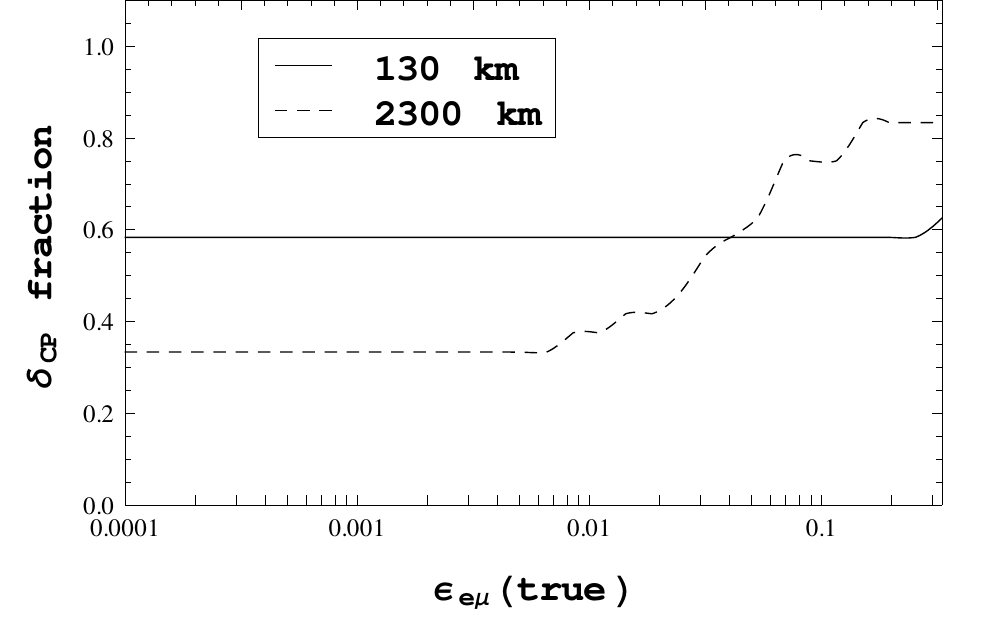}\\
\includegraphics[width=0.5\textwidth]{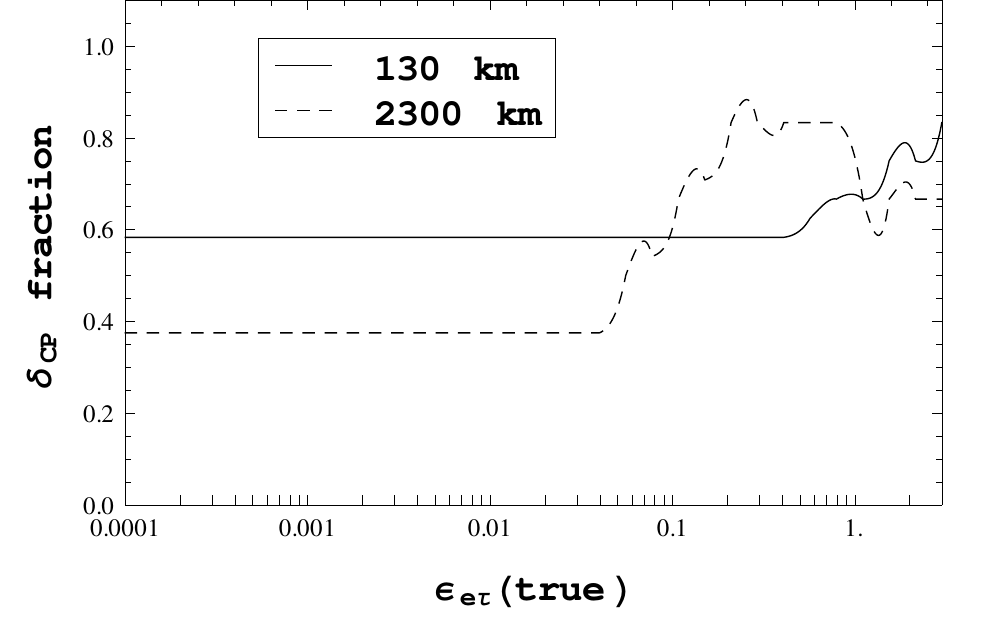}&
\includegraphics[width=0.5\textwidth]{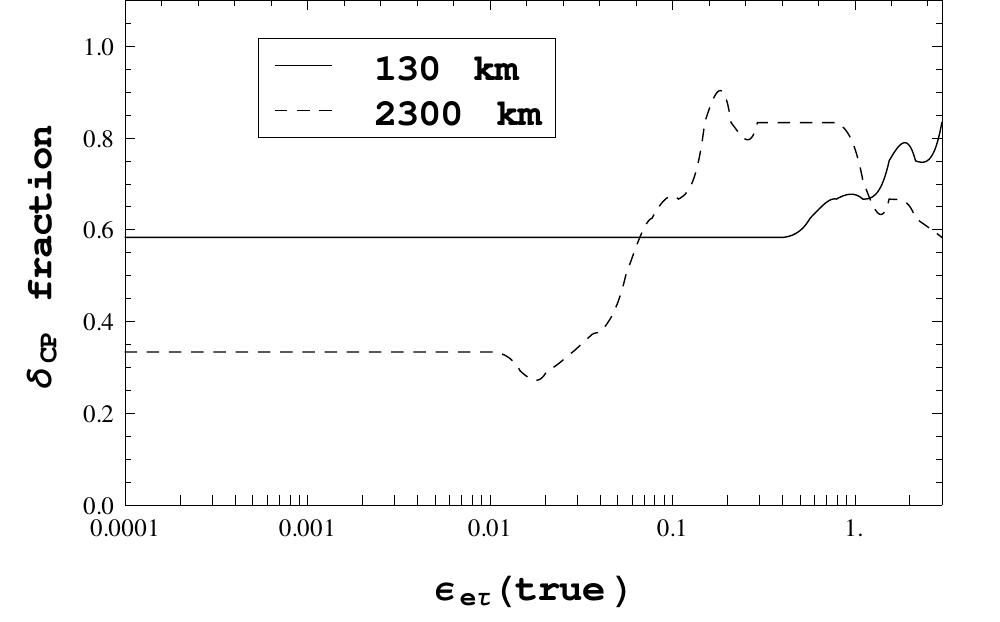}
\end{tabular}
\caption[] {{\small $\delta_{CP}$ fraction for two different baselines 130 Km and 2300 Km at $3\sigma$ considering NSIs $\ve_{ee}$, $\ve_{e\mu}$ and $\ve_{e\tau}$. The  left hand panel is for NH and the right hand panel is for IH.}}
\label{fig:delta2}
\end{figure}

In figure \ref{fig:delta1} we have shown the $\Delta \chi^2$ values versus $\delta (true)$  for SM interactions of neutrinos with matter from which the discovery 
reach can be obtained at different confidence levels. Particularly at $3 \sigma $ confidence level  for 130 Km baseline the $CP$ violation could be discovered 
around $\delta_{CP}$ fraction of 0.59 of the possible $\delta$ values for both normal and inverted hierarchy (true value) whereas for 2300 Km
baseline these values are about 0.37 for normal hierarchy and 0.35 for inverted hierarchy. Here we define $\delta_{CP}$ fraction as the fraction of the total alowed range (0 to 2$\pi$) for the $CP$ violating phase over which $CP$ violation can be discovered. So for longer baseline for normal hierarchy the discovery reach is better than the inverted hierarchy. The discovery reach for longer baseline of 2300 Km  was shown earlier by Coloma {\it et al}  \cite{others6}. So with only SM the short baseline like 130 Km seems to be better for good discovery reach of $CP$ violation. This was observed earlier by different authors \cite{Agarwalla:2011hh,others4,Coloma:2012ma,Coloma:2012ut,betabeam}. 
However, the short baseline may not be always better in presence of NSI which we discuss below.

\begin{figure}[H]
\centering
\begin{tabular}{cc}
\includegraphics[width=0.5\textwidth]{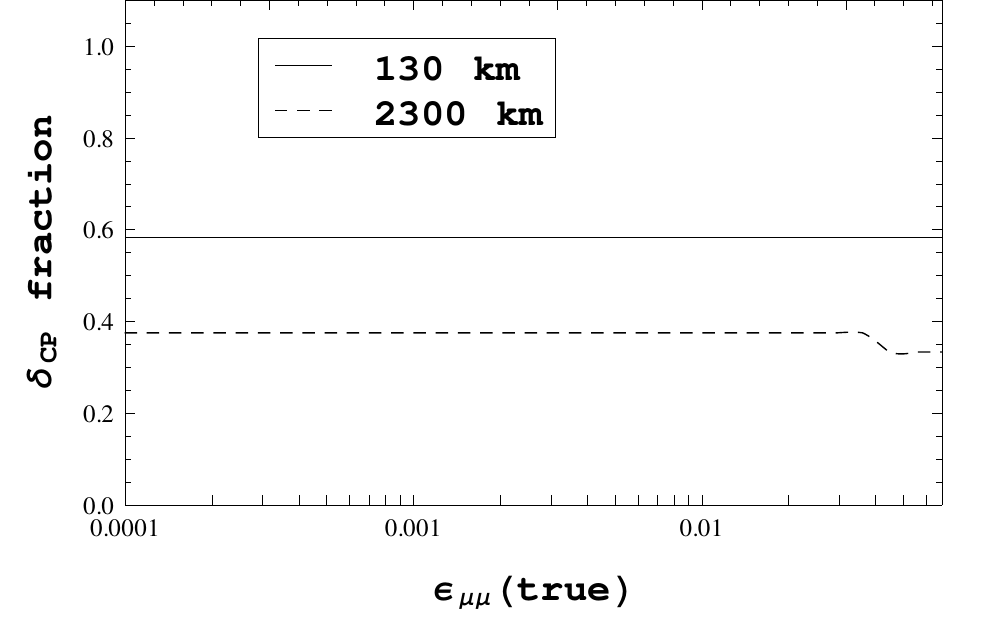}&
\includegraphics[width=0.5\textwidth]{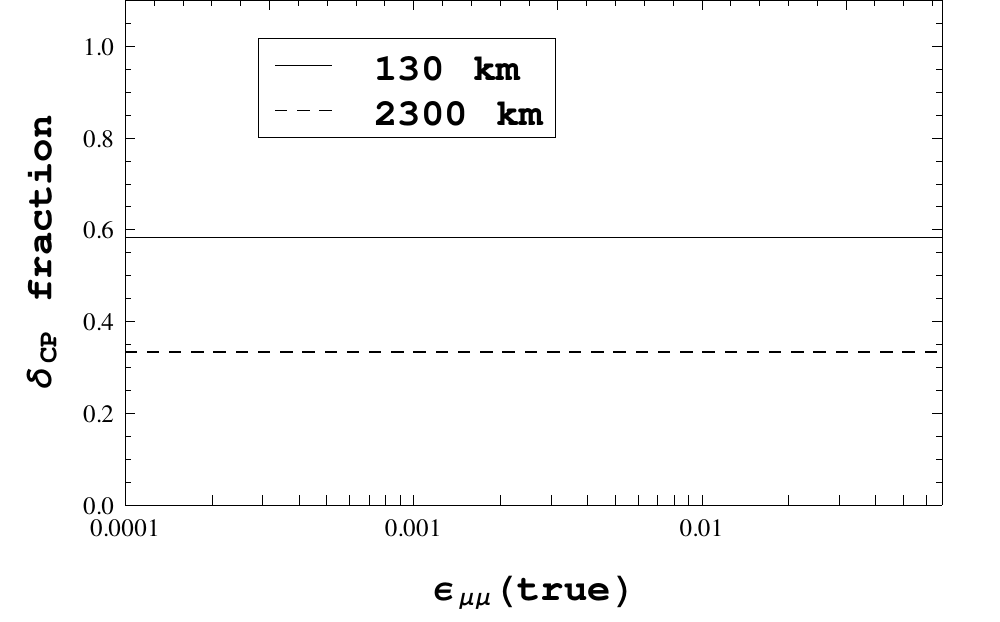}\\
\includegraphics[width=0.5\textwidth]{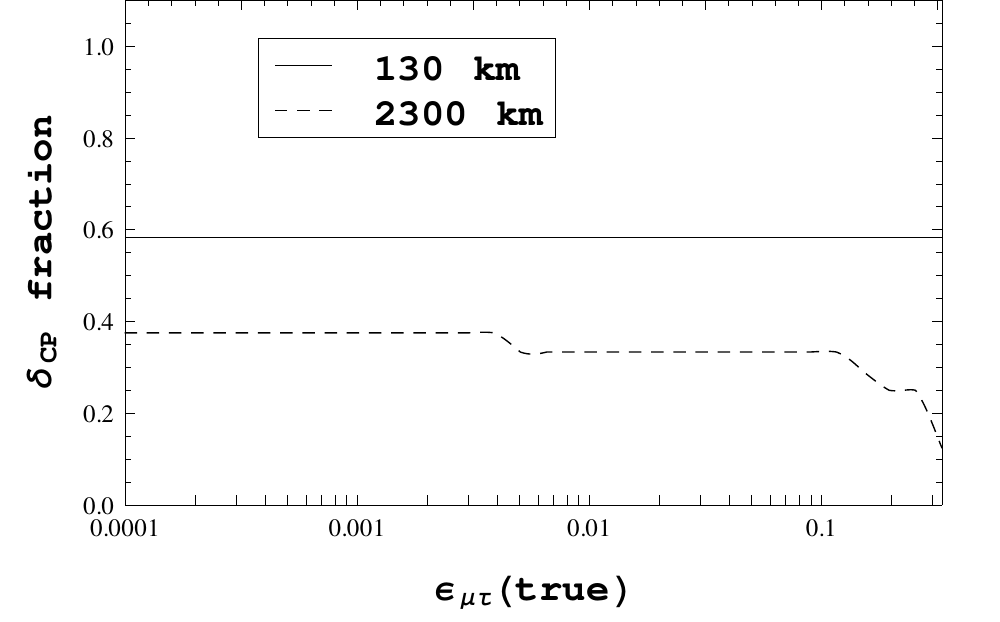}&
\includegraphics[width=0.5\textwidth]{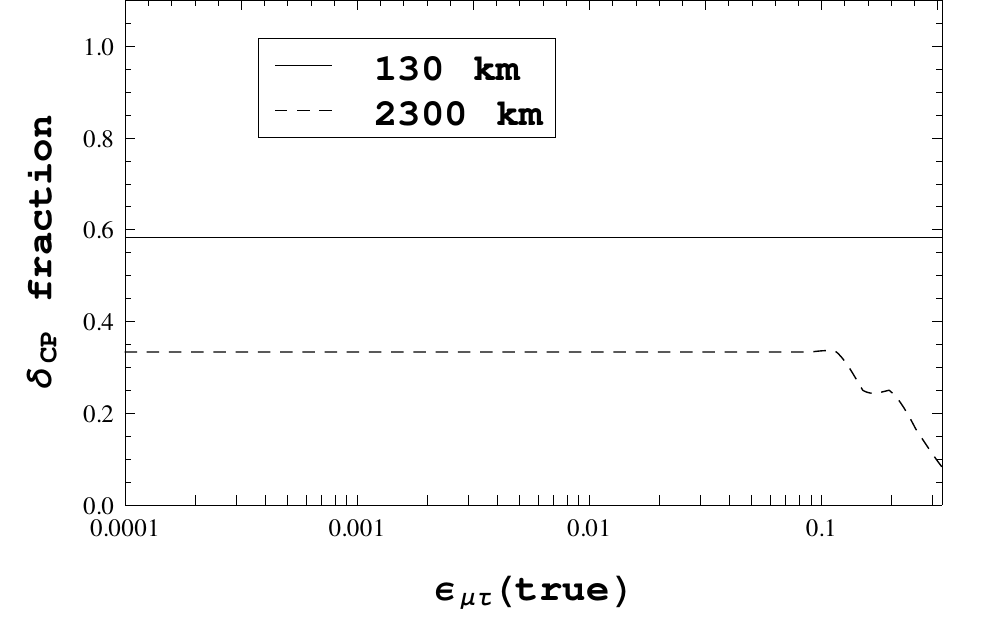}\\
\includegraphics[width=0.5\textwidth]{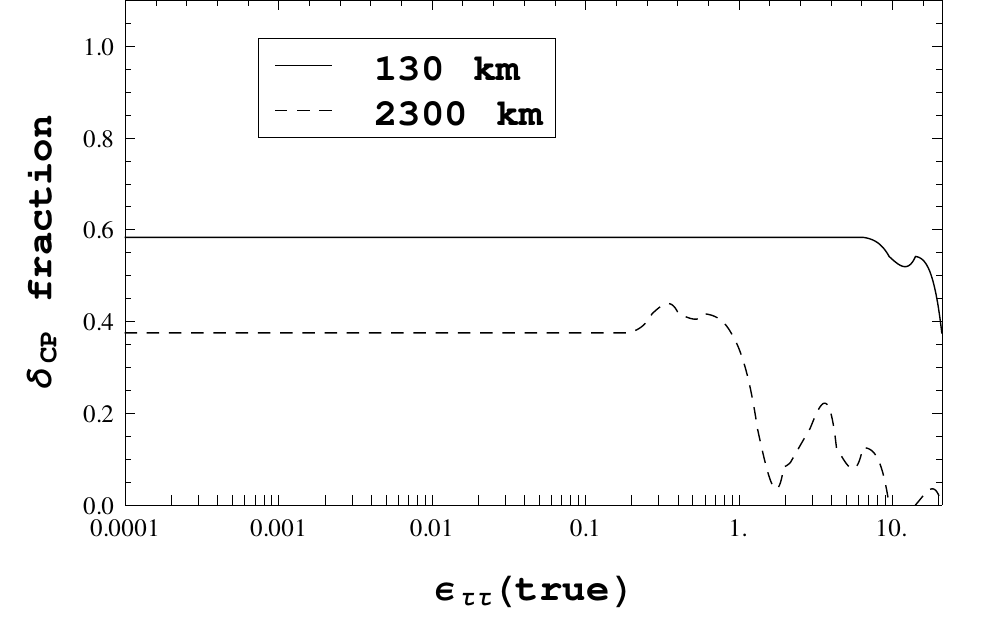}&
\includegraphics[width=0.5\textwidth]{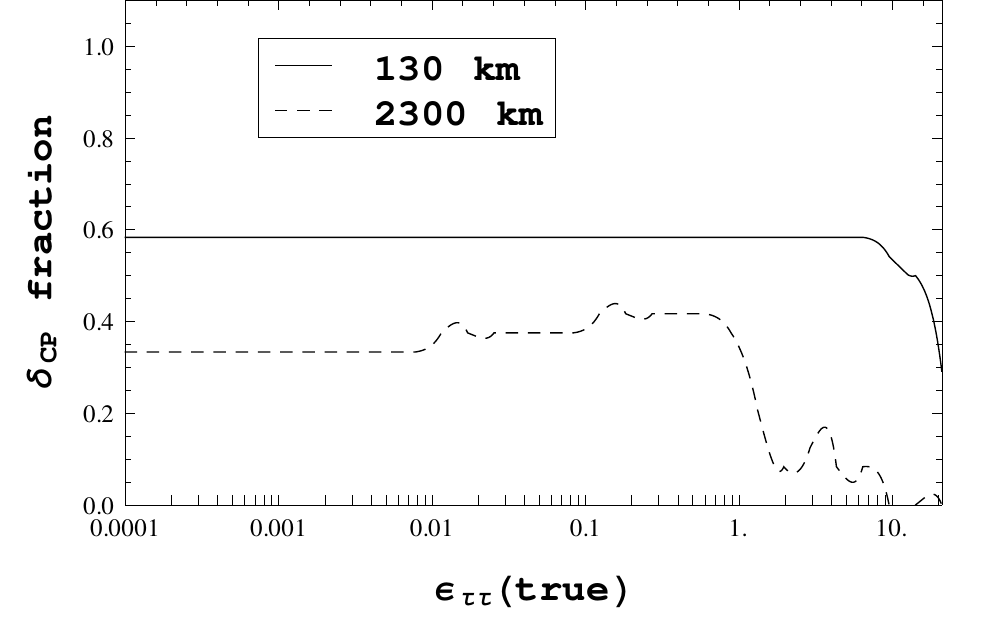}
\end{tabular}
\caption[] {{\small  $\delta_{CP}$ fraction for two different baselines 130 Km and 2300 Km at $3\sigma$ considering NSIs $\ve_{\mu\mu}$, $\ve_{\mu\tau}$ and $\ve_{\tau\tau}$. The  left hand panel is for NH and the right hand panel is for IH.}}
\label{fig:delta}
\end{figure}

\begin{figure}[H]
\centering
\begin{tabular}{cc}
\includegraphics[width=0.45\textwidth]{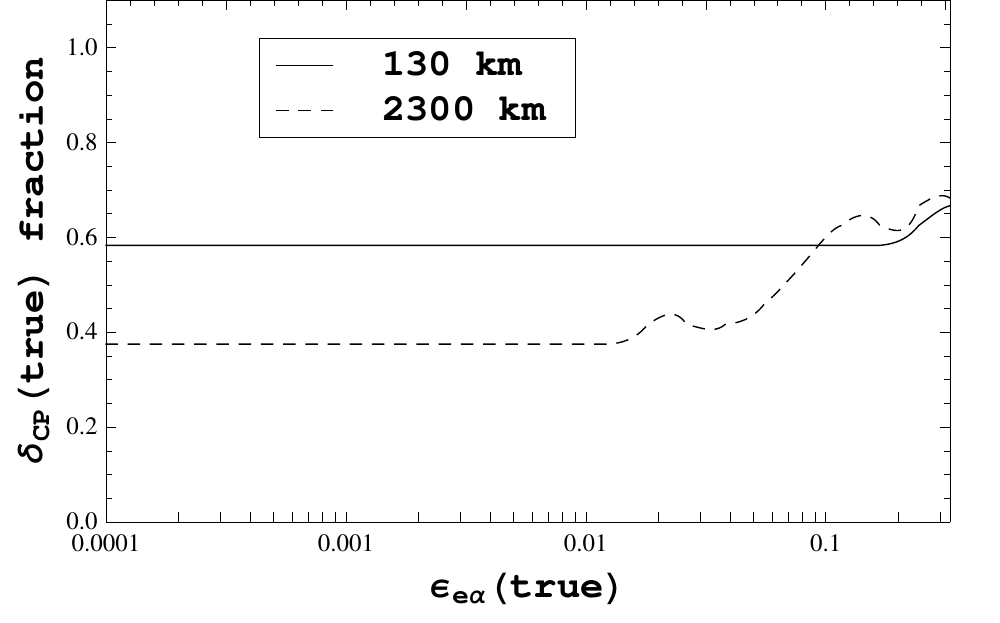}&
\includegraphics[width=0.45\textwidth]{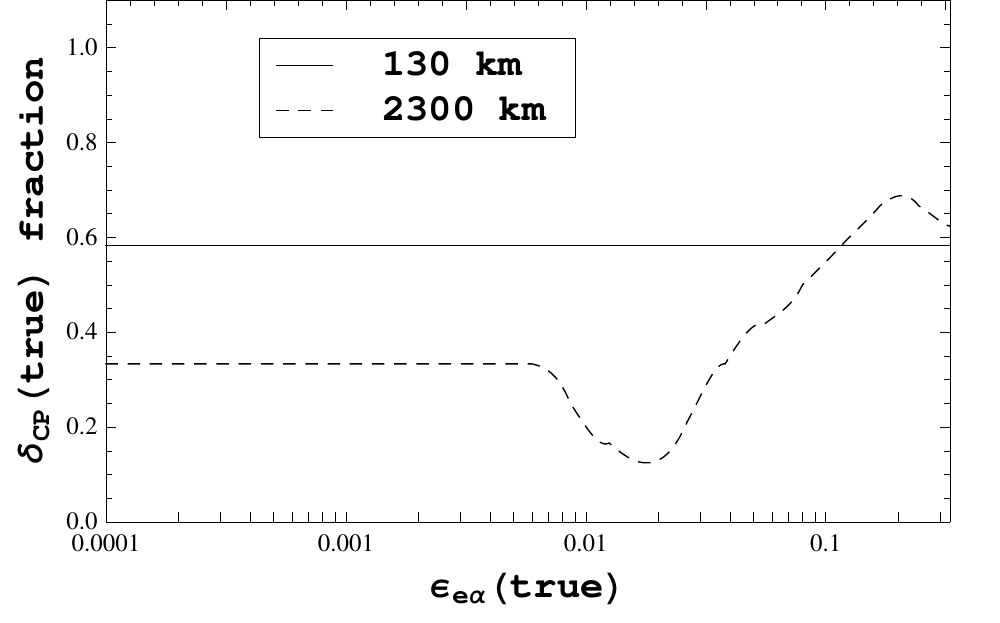}\\
\includegraphics[width=0.45\textwidth]{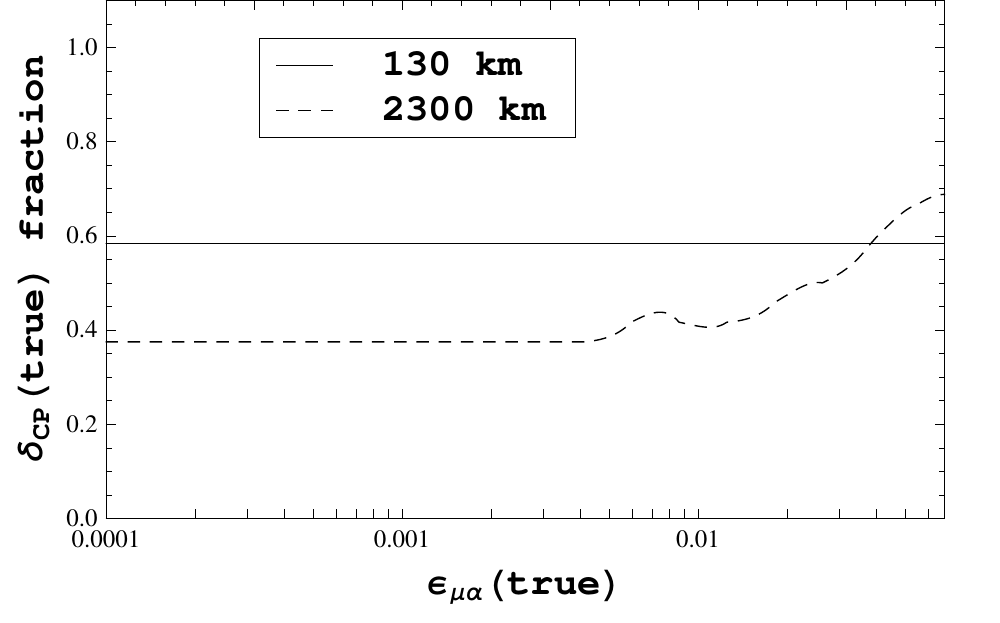}&
\includegraphics[width=0.45\textwidth]{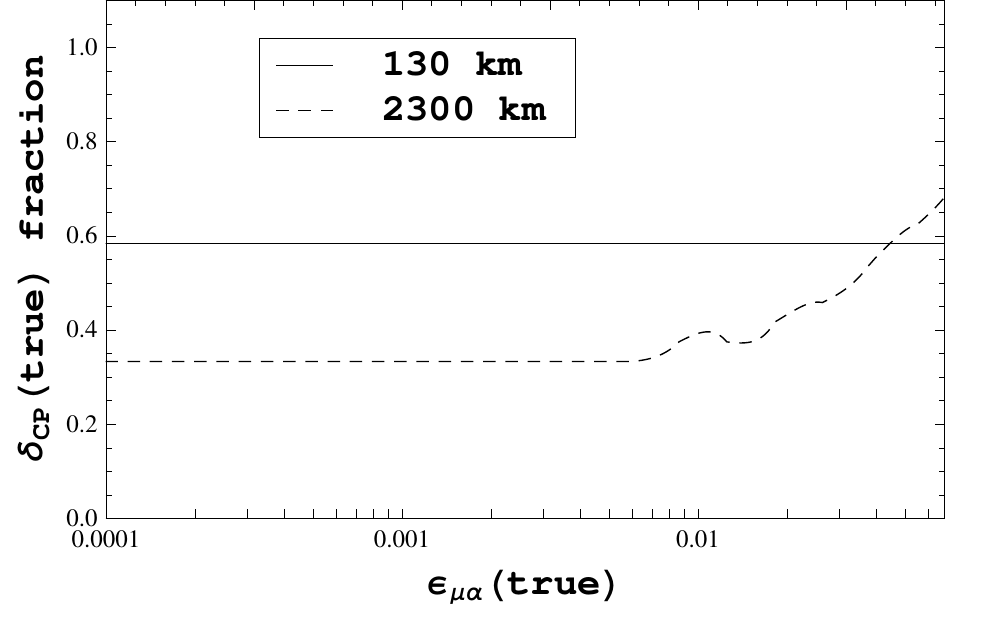}\\
\includegraphics[width=0.45\textwidth]{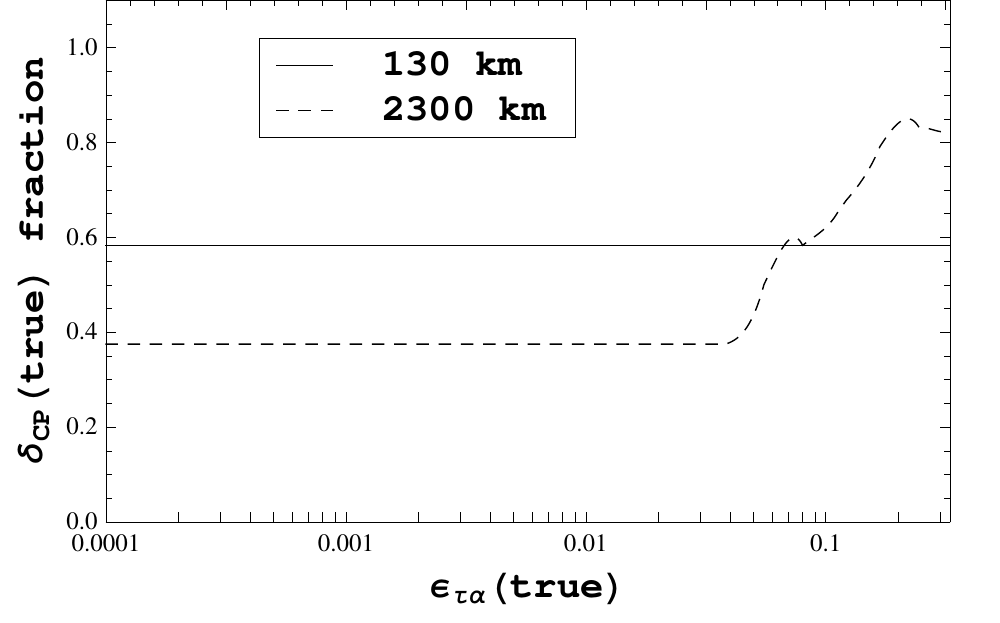}&
\includegraphics[width=0.45\textwidth]{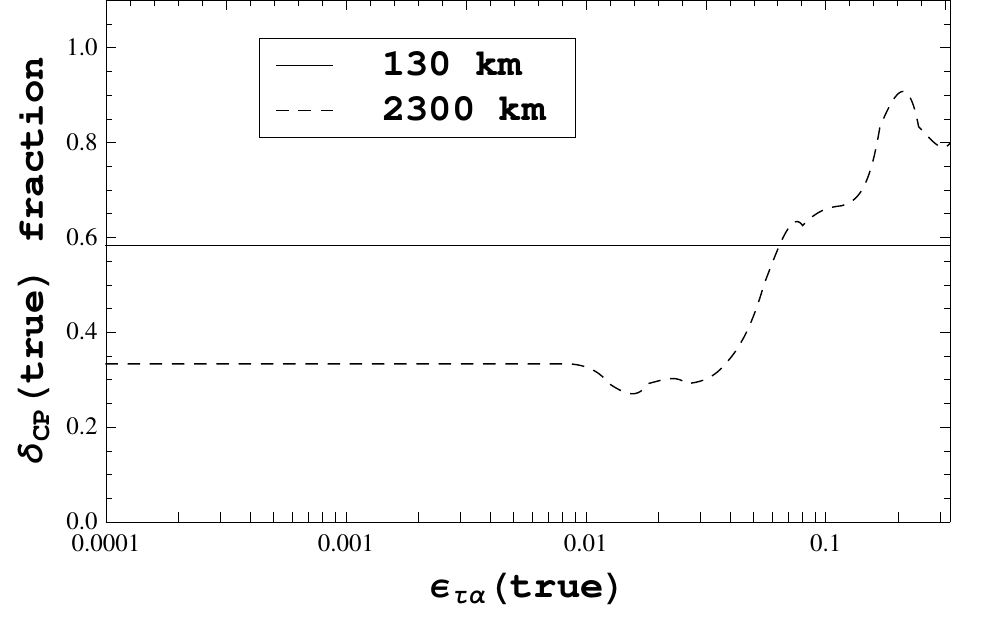}
\end{tabular}
\caption[] {{\small  $\delta_{CP}$ fraction for two different baselines 130 Km and 2300 Km at $3\sigma$ considering NSIs $\ve_{e \a}$, $\ve_{\mu\a}$ and $\ve_{\tau\a}$.The left panel is for NH and the right panel is for NH.}}
\label{fig:delta4}
\end{figure}

For finding the discovery reach of $CP$ violation due to $\delta$ in presence of NSIs' we have fixed $\delta^{test}$ to 0 and $\pi$ and have marginalized over hierarchy like the earlier case. We have considered same value for NSI$^{test}$ and NSI$^{true}$ while varying them. We have varied the value of NSIs' in the range of $10^{-4}$ to the upper bound of respective NSIs' provided by the Model independent bounds given in Table \ref{table:bound}. At $3 \sigma $ confidence level, we
show the $\delta_{CP}$ fraction for different NSI values.  We have considered one NSI at a time. In figure \ref{fig:delta2} we have compared the discovery reach of $CP$ violation for 130 Km baseline and 2300 Km baseline in presence of NSIs - $\ve_{ee}$, $\ve_{e\mu} $ and $\ve_{e\tau}$ for both the hierarchies.  
For  $\ve_{ee}$ in the range as mentioned above for 130 Km baseline seems to give a better discovery reach of $CP$ fraction for both the hierarchies. However, for 2300 Km baseline the $CP$ fraction starts  dropping for $\ve_{ee} \gtrsim $0.2 for NH and 0.1 for IH.  For  $\ve_{e\mu} $, 130 Km baseline has better discovery reach for both the neutrino mass hierarchies upto about $\ve_{e\mu} \lesssim $ 0.15 for NH and 0.25 for IH and for 2300 Km baseline it has better discovery reach for both the neutrino mass hierarchies for  $\ve_{e\mu} \gtrsim $ 0.004 for NH and 0.006 for IH. For $\ve_{e\tau}$ 130 Km baseline is better in the range 0.1 $\lesssim \ve_{e\tau} \lesssim$ 3.0 for NH and 0.4 $\lesssim \ve_{e\tau} \lesssim$ 3, for IH. For 2300 Km  baseline is better in the range 0.03 $\lesssim \ve_{e\tau} \lesssim$ 3.0 for NH and 0.02 $\lesssim \ve_{e\tau} \lesssim$ 3, for IH. For NSI of the order of $10^{-4}$ there is almost negligible effect on the $CP$ violation discovery reach and the  $CP$ fractions correspond to almost that for SM value.  One may note that in the expression of oscillation probability (following perturbation method) in presence of particularly two NSIs' $\ve_{e\mu}$  and $\ve_{e\tau}$ more $\delta $ dependent terms appear. This could be the reason for better discovery reach of $CP$ violation in presence of these two NSIs' in comparison to the case of only SM as seen in the figures.  In figure \ref{fig:delta} we have compared the discovery reach of $CP$ violation for 130 Km baseline and 2300 Km baseline in presence of NSIs -$\ve_{\mu\mu}$, $\ve_{\mu\tau}$ and $\ve_{\tau\tau}$. Here we see for the entire allowed range of these NSI parameters there is better discovery reach for 130 Km baseline. 
 For $\ve_{\mu\tau}$  for 130 Km baseline there is no significant deviation of $\delta_{CP}$ fraction from its' SM value. However, for 2300 Km baseline this fraction starts decreasing from its' SM value above around 0.004 for NH and above around 0.1 for IH. For $\ve_{\tau\tau}$  for  130 Km baseline this fraction starts deviating from its' SM value above around 9.0 for NH and above around 10.0 for IH.  For 2300 Km baseline this fraction starts deviating from its' SM value above around 0.2 for NH and above around 0.01 for IH. 
One may note that the effect of the NSI - $\ve_{e\mu}$, $\ve_{\mu\mu}$, $\ve_{\mu\tau}$ and $\ve_{\tau\tau}$  on the $CP$ violation discovery is negligible due to their relatively smaller values.  The straight line in the  figures \ref{fig:delta2} and \ref{fig:delta} corresponding to those NSIs' indicates this negligible effect. Particularly for the stringent bound of $\ve_{\mu\mu}$ the $CP$ fraction almost correspond to that for only SM case except for the higher value of this NSI around 0.25 for longer baseline with normal hierarchy there is departure from SM value. 

Next, in figure \ref{fig:delta4}  we consider each generation separately which corresponds to three NSIs' at a time. We consider the same values
for all three such NSIs' and consider their uppermost value satisfying their individual upper bounds.    
In this case we have fixed $\delta^{test}$ to 0 and $\pi$ and have marginalized over hierarchy. For three NSIs' corresponding to each generation we have considered same test and true values for all three NSIs'. For $\ve_{e\a}$  for  130 Km baseline this fraction starts deviating from its' SM value above around 0.25 for NH and  for IH it is around SM value. For 2300 Km baseline this fraction starts deviating from its' SM value above around 0.015  for NH and above around 0.006 for IH. For $\ve_{\mu\a}$  for  130 Km baseline this fraction remains around SM value  over the considered range for both hierarchies.   For 2300 Km baseline this fraction starts deviating from its' SM value above around 0.005 for NH and above around 0.008 for IH. For $\ve_{\tau\a}$  for   130 Km baseline this fraction remains around SM value  over the considered range for both hierarchies.   For 2300 Km baseline this fraction starts deviating from its' SM value above around 0.5 for NH and above around 0.01 for IH.

\subsection{Discovery reach of $CP$ violation due to both $\delta$ and complex NSI phases }
\begin{figure}[H]
\centering
\begin{tabular}{cc}
\includegraphics[width=0.45\textwidth]{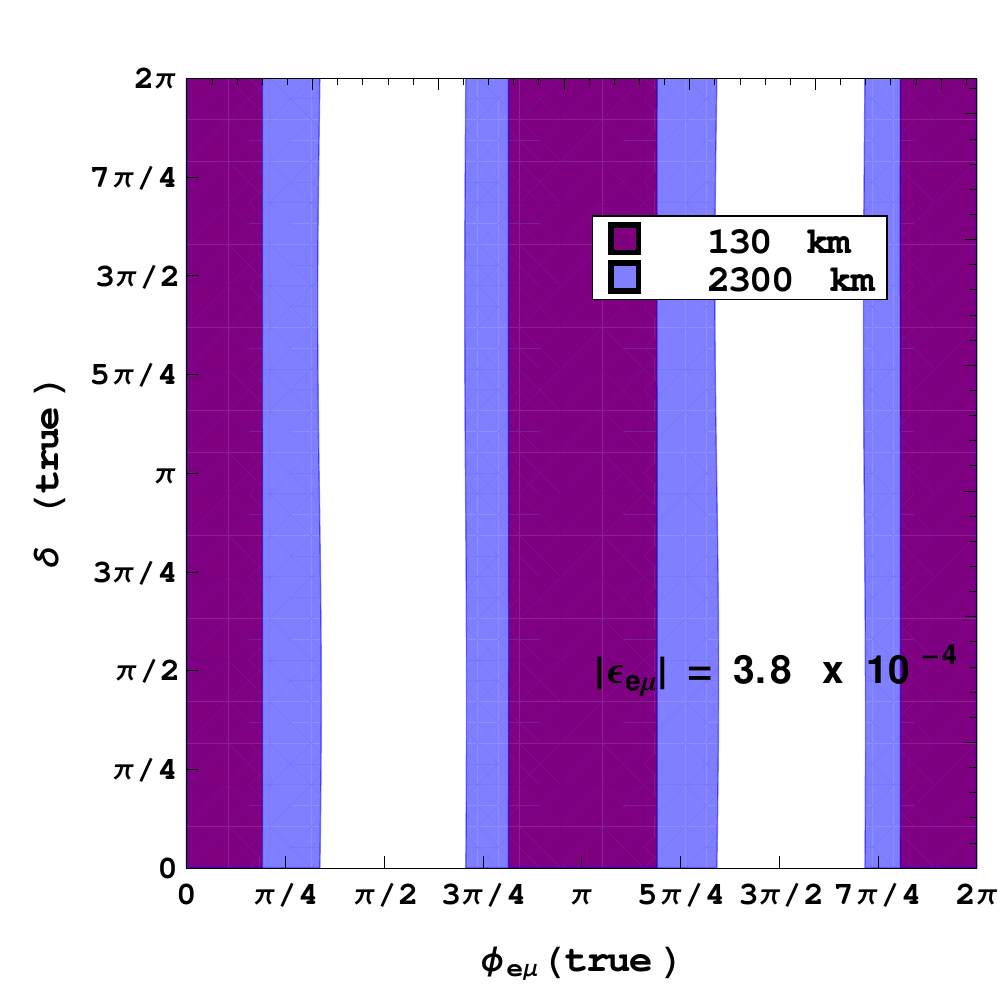}&
\includegraphics[width=0.45\textwidth]{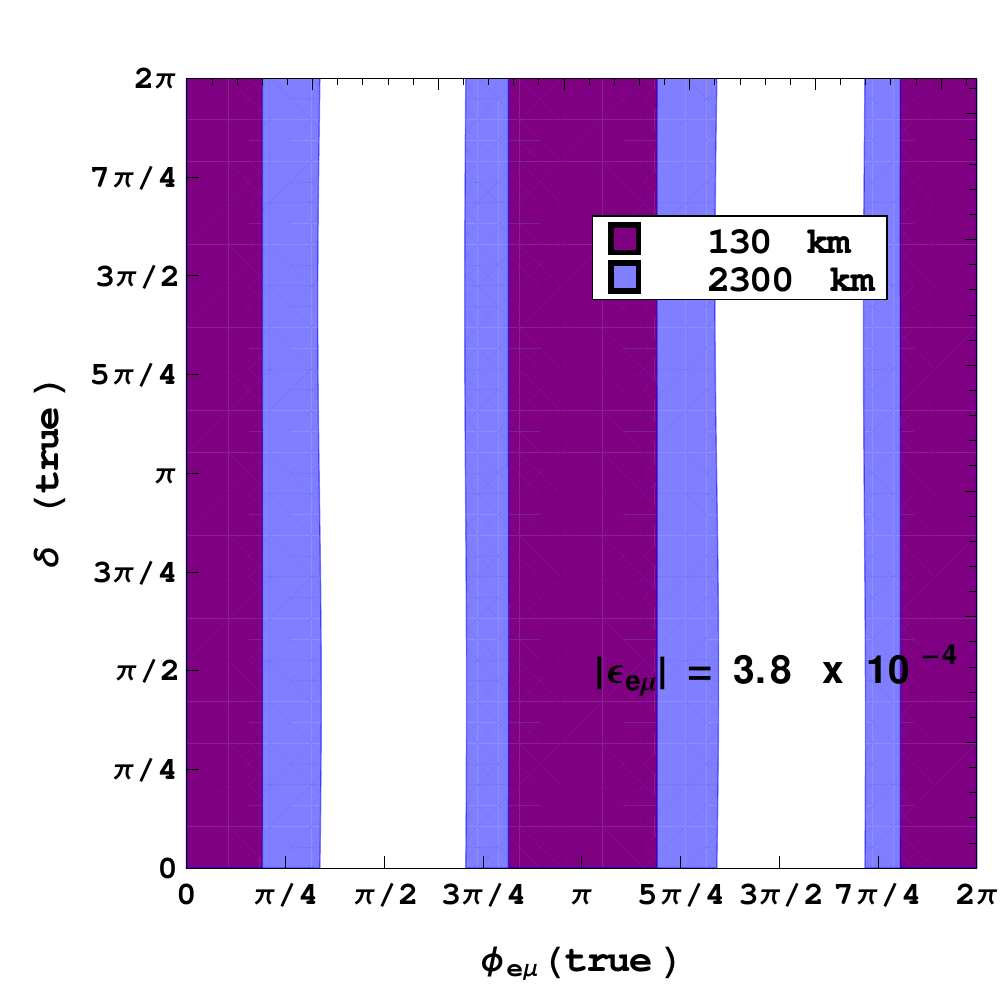}\\
\includegraphics[width=0.45\textwidth]{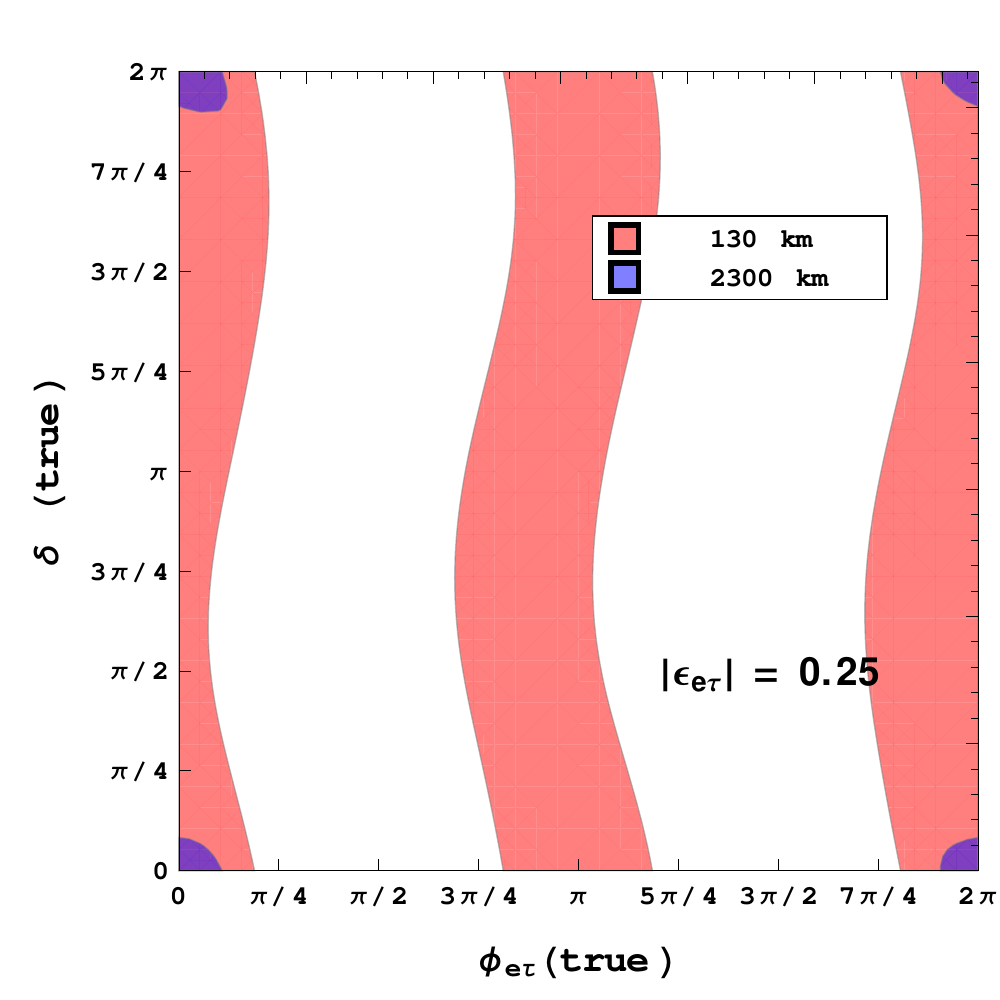}&
\includegraphics[width=0.45\textwidth]{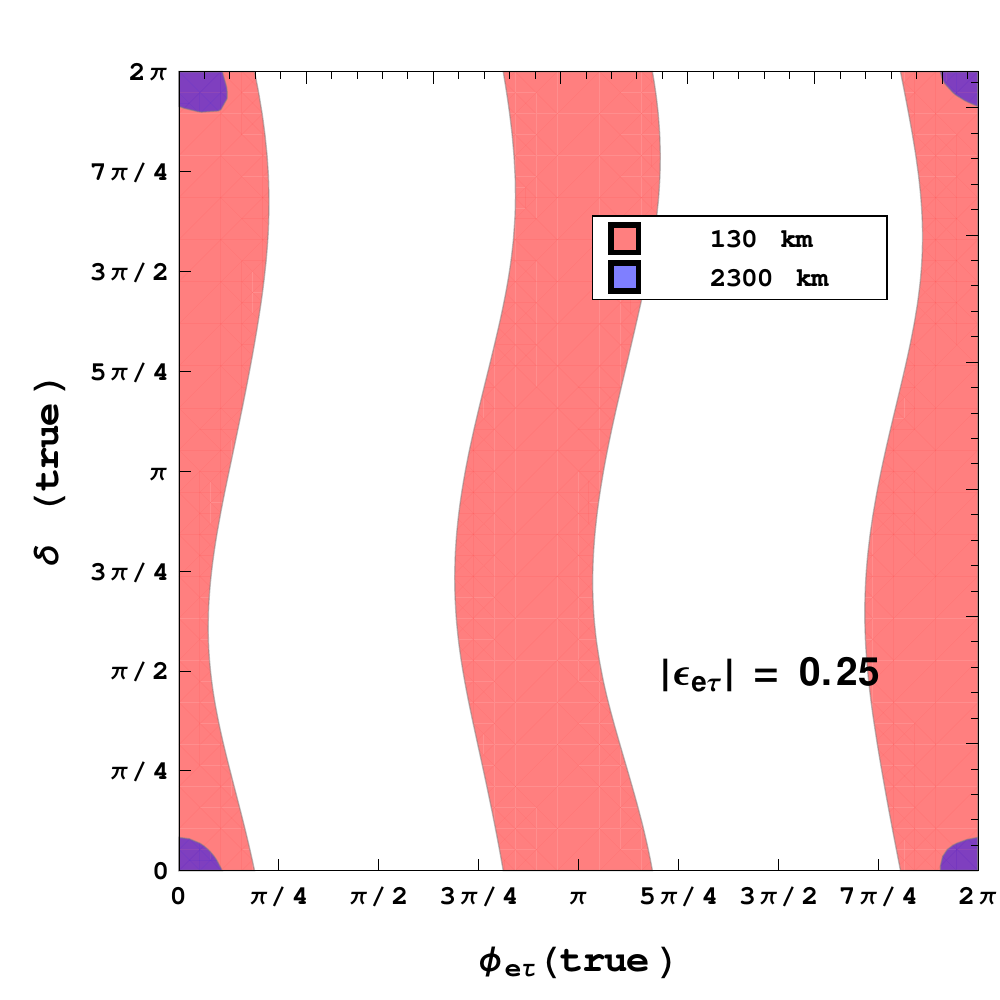}\\
\includegraphics[width=0.45\textwidth]{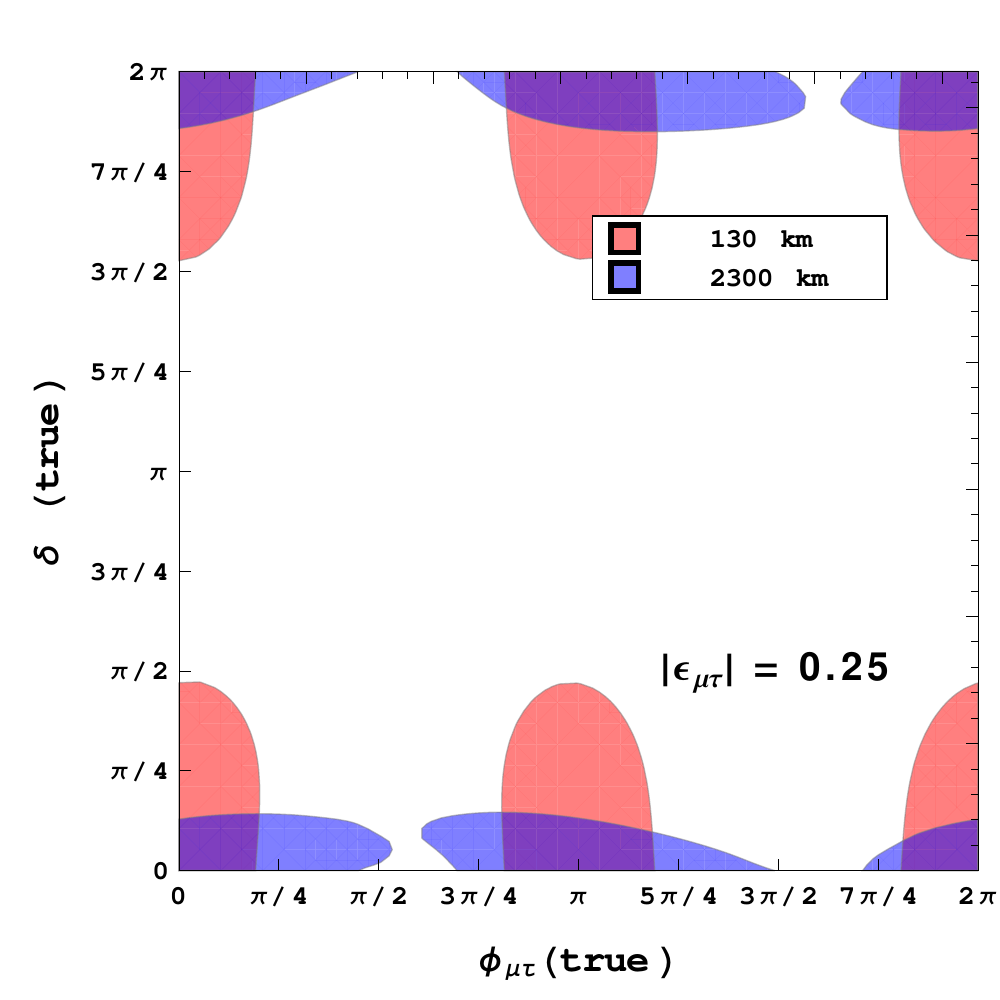}&
\includegraphics[width=0.45\textwidth]{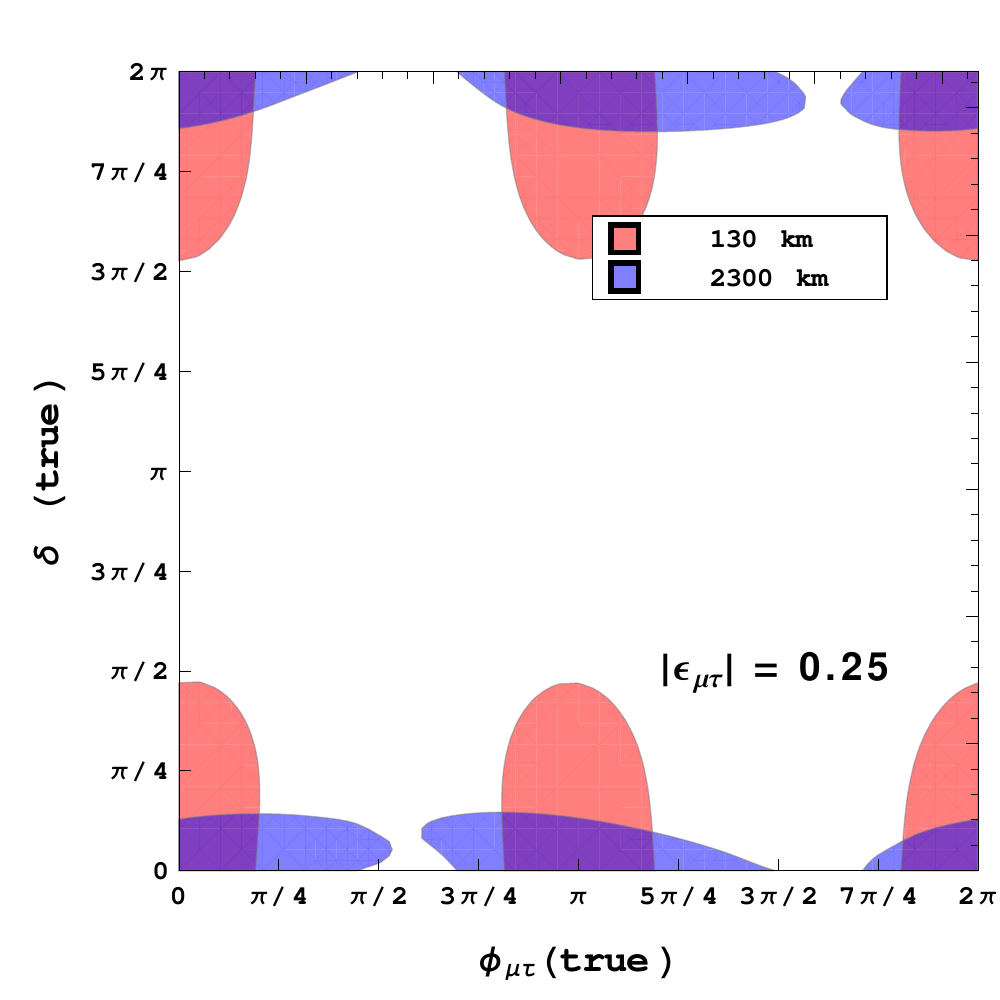}
\end{tabular}
\caption[] {{\small  Allowed region (unshaded) for total $CP$ violation discovery reach $\delta_{CP}$ fraction  for two different baselines 130 Km and 2300 Km at $3\sigma$ considering NSI phases $\phi_{e\mu}$, $\phi_{e\tau}$ and $\phi_{\mu\tau}$. The left panel is for NH and the right panel is for IH.}}
\label{fig:nsiphase2}
\end{figure}

Here we show the contours showing discovery reach of  $CP$ violation due to both $\delta$ and complex NSI phases for both the baselines 130 Km and 2300 Km  baselines. The off-diagonal 
elements in the NSI matrix - $\ve_{e\mu}$, $\ve_{e\tau}$ and $\ve_{\mu\tau}$ could be complex whereas other NSIs' are real. For that we have fixed $\delta^{test}$ \& $\phi^{test}_{NSI}$ to 0 and $\pi$ and then have marginalized over hierarchy (test) over all of them for every pair
of $\delta^{true}$  and $\phi^{true}_{NSI}$ considering one NSI at a time. We have fixed absolute value of the respective 
NSIs to their uppermost allowed values given by the Model dependent bounds as shown in table \ref{table:bound}. In figure \ref{fig:nsiphase2} we have shown the allowed discovery region in the plane of Dirac phase $\delta $ and one of the three NSI phases (true)  to show the discovery reach of total $CP$ violation due to both $\delta$ and one of the NSI phases. In these plots the unshaded regions corresponds to the total discovery reach of $CP$ violation. For $\delta =0, \pi, 2 \pi$ and the NSI phases also having those values obviously one can not get $CP$ violation discovery. Corresponding to $\delta $ values very near to $\delta = \pi$ with NSI phases having one of those $CP$ conserving values, sometimes the region for which 
$CP$ violation can not be discovered, is too small to be seen in the figures. For NSI phase $\phi_{e\mu}$ there are some values of it for which $CP$ violation discovery reach is not possible for any value of $\delta $. For longer and shorter baseline as the pattern is similar 
for no observation of $CP$ violation for any $\delta$ values so this problem may not be solved by considering combination of two baselines with one shorter and another longer baseline. For 130 Km baseline these values are from 0 to $3 \pi/16$; $ 13 \pi/16  $ to $19 \pi/16$ and $29 \pi/16$ to $2 \pi$. For 2300 Km baseline  these values are from 0 to $5 \pi/16$; $ 23 \pi/32  $ to $4 \pi/3$ and $55 \pi/32$ to $2 \pi$. This feature is due to highly stringent constraint on  $|\varepsilon_{e\mu}|$ which is somewhat smaller than other NSIs'. For
further smaller values of $|\varepsilon_{e\mu}|$ the possible discovery region will be further reduced. 
However, unlike $\phi_{e\mu}$ for $\phi_{e\tau}$ there is better discovery reach for longer baseline almost covering the entire region. One may note here that even for real  $|\varepsilon_{e\tau}|$ there is better discovery reach of $\delta_{CP}$ fraction for higher value of this NSI as can bee seen in figure \ref{fig:delta2}.

\section{Conclusion}
We have studied the possible $CP$ violation discovery reach due to Dirac phase $\delta$ in the leptonic sector through neutrino oscillation experiments with superbeam as neutrino source.  
To study the $CP$ violation discovery reach in presence of NSIs'  we have considered two experimental set-ups - one with a long baseline of 2300 Km directed towards a 100 Kt Liquid argon detector and the other with a relatively shorter baseline of 130 Km directed towards a 500 Kt Water Cherenkov detector  
and  have considered   central values of $\theta_{13}$ with errors as shown in table \ref{table:mix}  coming from recent reactor experiments for our analysis.

In figures  \ref{fig:probEL1},  \ref{fig:probEL2}, \ref{fig:probEL3} we have shown  
the NSI effect $\Delta_p$ in the particular oscillation channel $\nu_{\mu}\rightarrow \nu_e$ (which is the most important channel for
$CP$ violation discovery) for different baseline length $L$ and different neutrino energy $E$. It is found that
oscillation probability $P_{\nu_\mu \rightarrow \nu_e}$ changes significantly in presence of NSI $\varepsilon_{e\mu} $ and  $\varepsilon_{e\tau}$ in comparison to other  NSIs for longer baselines and higher energies. It is known that in presence of SM interactions of neutrinos with matter the shorter baseline can give better discovery reach for $CP$ violation \cite{Agarwalla:2011hh,others4,Coloma:2012ma,Coloma:2012ut,betabeam}. However, as we have shown that in presence of NSI this may not be always true. Particularly, for NSIs' $\varepsilon_{e\mu} $ and  $\varepsilon_{e\tau}$ (considering real) the $CP$ violation discovery reach could be better than that for SM  case for longer baseline like 2300 Km at above some values of NSI as shown in figures \ref{fig:delta2}. Considering presence of three NSIs' at a time with their same values corresponding to a particular generation 
we find the similar observations for all three cases as shown in \ref{fig:delta4}. But is is very important to note here that this better discovery reach in presence of NSI is only possible provided that NSI has already been discovered. If NSI is not discovered then the
shorter baseline is preferred for $CP$ violation discovery to reduce the effect of real NSI.

NSIs' - $\ve_{e\mu}$, $\ve_{e\tau}$ and $\ve_{\mu\tau}$ could be complex. We have considered the corresponding phases 
$\phi_{e\mu}$, $\phi_{e\tau}$ and $\phi_{\mu\tau}$ respectively in the analysis of the discovery reach of total $CP$ violation. The $CP$ violation discovery region has been shown in the $\phi_{ij} - \delta $ plane 
in figure \ref{fig:nsiphase2}. In presence of $\phi_{e\mu}$ the shorter baseline is found to be better for total $CP$ violation discovery reach. But for $\phi_{e\tau}$  the longer baseline is consistently better. However, for $\phi_{\mu\tau}$ with its  certain combinations with 
Dirac phase $\delta$ sometimes shorter and sometimes longer baseline seems to be better. Due to this feature particularly for total $CP$ violation discovery reach in presence of NSI phase $\phi_{\mu\tau}$ the combination of two baselines might give better discovery reach.   
One may note here for complex NSI even if $|\mbox{NSI}|$ is known still  there could be 
some values of NSI phases for NSI $\ve_{e\mu}$ as seen in figure \ref{fig:nsiphase2} for which neither the $CP$ violation nor any values of $\delta$ could be discovered. Even for lesser NSI values than that considered in the figures, this kind of feature is expected.  So in presence of NSI it could turn out that non-observation of $CP$ violation might not imply the absence of Dirac phase
$\delta $ which could have been observable in presence of SM interactions only for neutrinos with matter.

In this paper we have not discussed about the discovery reach of NSI  as based on our discussion in section III the two experimental set-up which we have chosen are not really optimized for doing such analysis. NSI effects are found to be more in further longer baselines as shown in figures \ref{fig:probEL1}, \ref{fig:probEL2} and \ref{fig:probEL3} which we have not considered in this work. Considering the possibility of 
the presence of NSIs in nature it seems sometimes combination of both short and long baseline might improve the $CP$ violation discovery reach  in
the leptonic sector through neutrino oscillation experiment with superbeam facility but this strategy may not work for complex $\ve_{e\mu}$.

\hspace*{\fill}
\newpage
\noindent
{\bf  Acknowledgment:} AD thanks Council of Scientific and Industrial Research, India for financial support through
Senior Research Fellowship (EMR No. 09/466(0125)/2010-EMR-I) and ZR  thanks University Grants Commission, Govt. of India for providing research fellowships. AD thanks S. K. Raut for discussion on GLoBES.
We thank L. Whitehead  and Luca Agostino for providing  migration matrices for Liquid Argon detector and large scale water Cherenkov detector (as studied by MEMPHYS collaboration) respectively and for their other helpful communications.

\end{document}